    \documentclass{aa}  

\usepackage{txfonts}
\usepackage{graphicx,lipsum,overpic}
\usepackage{subcaption}
\usepackage{amssymb,amsmath}
\usepackage[version=4]{mhchem}

\newcommand{\ignore}[1]{}

\def\aap {{A\&A}}

\def\apjl {{ApJL}}

\def\apjs {{ApJS}}

\begin{document}

   \title{Connecting planet formation and astrochemistry }

    \subtitle{C/O and N/O of warm giant planets and Jupiter-analogs}

   \author{Alex J. Cridland\inst{1}\thanks{cridland@strw.leidenuniv.nl}, Ewine F. van Dishoeck\inst{1,2}, Matthew Alessi\inst{3}, \& Ralph E. Pudritz\inst{3,4}
          }

   \institute{
   $^1$Leiden Observatory, Leiden University, 2300 RA Leiden, the Netherlands \\ $^2$Max-Planck-Institut f\"ur Extraterrestrishe Physik, Gie{\ss}enbachstrasse 1, 85748 Garching, Germany \\ $^{3}$Department of Physics and Astronomy, McMaster University, Hamilton, Ontario, Canada, L8S 4E8 \\ $^4$Origins Institute, McMaster University, Hamilton, Ontario, Canada, L8S 4E8
             }

   \date{Received \today}


  \abstract
  {
The chemical composition of planetary atmospheres has long been thought to store information regarding where and when a planet accretes its material. Predicting this chemical composition theoretically is a crucial step in linking observational studies to the underlying physics that govern planet formation. As a follow-up to a study of hot Jupiters in our previous work, we present a population of warm Jupiters (semi-major axis between 0.5-4 AU) extracted from the same planetesimal formation population synthesis model as used in our previous work. We compute the astrochemical evolution of the protoplanetary disks included in this population to predict the carbon-to-oxygen (C/O) and nitrogen-to-oxygen (N/O) ratio evolution of the disk gas, ice, and refractory sources, the accretion of which greatly impacts the resulting C/O and N/O in the atmosphere of giant planets. We confirm that the main sequence (between accreted solid mass and atmospheric C/O) we found previously is largely reproduced by the presented population of synthetic warm Jupiters. And as a result, the majority of the population fall along the empirically derived mass-metallicity relation when the natal disk has solar or lower metallicity. Planets forming from disks with high metallicity ([Fe/H] $>$ 0.1) result in more scatter in chemical properties which could explain some of the scatter found in the mass-metallicity relation. Combining predicted C/O and N/O ratios shows that Jupiter does not fall among our population of synthetic planets, suggesting that it likely did not form in the inner 5 AU of the solar system before proceeding into a Grand Tack. This result is consistent with recent analysis of the chemical composition of Jupiter's atmosphere which suggests that it accreted most of its heavy element abundance farther than tens of AU away from the Sun. Finally we explore the impact of different carbon refractory erosion models, including the location of the carbon erosion front. Shifting the erosion front has a major impact on the resulting C/O ratio of Jupiter and Neptune-like planets, but warm Saturns see a smaller shift in C/O, since their carbon and oxygen abundances are equally impacted by gas and refractory accretion.
}


    \maketitle
%

\section{Introduction}

It is now well established that the study of an exoplanetary atmospheric carbon-to-oxygen ratio (C/O) represents an important step in understanding the physical processes that govern planet formation \citep{Oberg11,Helling14,Madu2014,Crid16a,Crid19b}. To date, measurements of atmospheric C/O have largely been carried out for hot Jupiters and hot Neptunes because their proximity to their host star make high signal to noise transmission and emission spectra more easily attainable \citep{Madu2012,Moses13,Brogi2014,Line2014,Brewer2016b,Gandhi2018,Pinhas2019,MacDonald2019}.

Farther away from their host star are cold Jupiters (a.k.a directly imaged planets), with orbital radii $\geq 8$ AU, that can be chemically characterized through the efforts of direct spectroscopy and interferometry. The GRAVITY consortium with their recent effort for $\beta$ Pic b \citep[at 9.2 AU,][]{GRAVITY2020}, have provided a precise measurement of C/O for that planet and shown that such a measurement is feasible with the interferometric mode of the Very Large Telescope \citep[VLTI,also see][]{GRAVITY2019}. This method of chemical characterization will compliment the efforts of the directly imaging community which have planned both Early Release Science\footnote{see:  \url{http://www.stsci.edu/jwst/observing-programs/approved-ers-programs/program-1386}} and Guaranteed Time Observations\footnote{see: \url{https://www.cosmos.esa.int/web/jwst-nirspec-gto/ifs-of-an-exoplanet-system}} with the James Webb Space Telescope (JWST) for planets at larger distances. 

At orbital radii between the hot and cold Jupiters are a population of exoplanets that have not been well studied chemically. These `warm' Jupiters are defined as having orbital radii between 0.5 - 10 AU. They orbit too close to their host star to be detectable by direct imaging, but far enough away that their detection via the transit method would be limited due to their long orbital period. With this definition Jupiter and Saturn, with effective temperatures of 134 K and 97 K respectively \citep{Aumann1969}, are classified as `warm' Jupiters. We note that this classification is not based on the effective temperature of the planet (which can depend strongly on internal processes), but instead only depends on the planet's orbital radius. 

\begin{figure}
    \centering
    \begin{overpic}[width=0.5\textwidth]{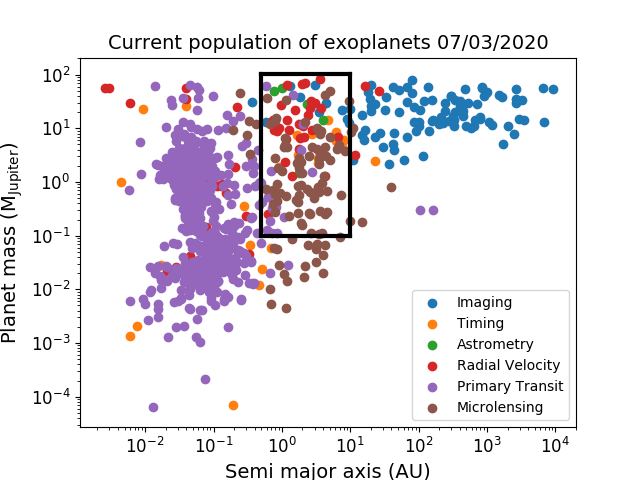}
        \put(45,25){\rotatebox{37}{\large Warm Jupiters}}
    \end{overpic}
    \caption{The current population of confirmed exoplanets extracted from http://exoplanet.eu/ on 07/03/2020. We note the range of mass and semi-major axis that define our warm Jupiters with black lines. The majority of planets in this range were discovered through direct imaging, radial velocity, and microlensing.}
    \label{fig:mass_a}
\end{figure}

In Figure \ref{fig:mass_a} we show the population of known exoplanets coloured by their primary discovery method\footnote{Extracted from exoplanet.eu}. Additionally we outline the mass and orbital radius range that is relevant for warm Jupiters with black lines. The majority of these planets were discovered through the radial velocity and microlensing techniques, with a few being discovered through primary transits, astrometry, and direct imaging. As already mentioned, these planets exist in a region of parameter space that make their chemical characterization difficult, and as such there are few examples where such a measurement has been attempted. Regardless, occurrence rate studies of giant planets have shown that Jupiter-analogs (giant planets orbiting between 3-6 AU) should be more abundant \citep[$\sim$ 3.3\% of stars,][]{Wittenmyer2011} than hot Jupiters \citep[$\sim$ 1.2\% of stars,][]{Wright2012}. This conclusion, which is supported by population synthesis studies \citep{Mordasini2009b,HP13,Chambers2018}, motivates our study of warm Jupiters.

To what extent can the chemical properties of Jupiter and Saturn be understood in the context of the warm Jupiters? While the formation of Jupiter and Saturn are still open questions, there are two leading formation scenarios that have been studied in the literature: planetesimal accretion near the water ice line \citep{Pollack1996,Alibert2005,Helled2014}, and pebble accretion \citep{Bitsch2015,Bosman2019}. Comparing the chemical properties of our modelled warm Jupiters to Jupiter and Saturn can help to differentiate between these different formation pathways. Another popular planet formation scenario is gravitational instability, which is thought to lead to planets on wider orbits than our giant planets \citep[see for example][]{DR2009}.

If Jupiter and Saturn formed through planetesimal accretion near the water ice line, then they would have to undergo a Grand Tack \citep{Walsh2011} to migrate out to their current orbital radius (from 1-3 AU to 5.5 and 9.5 AU respectively). This process, however, is very sensitive to the mass ratio of the two planets and requires particular orbital radii arrangement to function \citep{Raymond2014,Chametla2020}. In this way, there could be many solar systems in the galaxy that have planets that underwent similar formation histories to Jupiter, but did not undergo a Grand Tack. Our simulated population of warm Jupiters orbit at radii inward of 4 AU (see below), and hence can be thought of as Jupiter- and Saturn- analogs that did not undergo a Grand Tack. 

This work is a follow up to our previous work that studied the chemistry of a population of hot Jupiters \citep[][Paper 1]{Crid19c}. The population we study here is extracted from the same population synthesis model as was our hot Jupiter model in Paper 1 \citep[taken from][]{APC2020}. In Paper 1 we found a relation between the atmospheric C/O in these hot Jupiters to the fraction of their total mass that was accreted as solids. We dubbed this relation a `main sequence' of atmospheric C/O and highlighted the fact that solid accretion - as planetesimals in our model - are important for determining the bulk chemical properties of hot Jupiter atmospheres. The well known (empirically derived) mass-metallicity relation \citep{Kre14} directly follows from this main sequence. Its prediction - that higher mass planets have lower bulk metallicity - is explained by our main sequence as being caused by the fact that high mass planets tend to be more dominated by gas accretion than solid accretion.

Does this main sequence - and hence the mass-metallicity relation - continue to work for warm Jupiters? And can the chemical structure of Jupiter's atmospheres (and by extension its formation history) be explained by our planetesimal accretion model? Unlike hot Jupiters, the orbital radii of warm Jupiters ranges across large chemical gradients in the disk, including the water ice line (between 2-4 AU) and the carbon erosion front ($\sim$ 5 AU). As such, we expect to find a larger spread in atmospheric chemical compositions than we did in Paper 1. On the other hand, hot Jupiters are expected to have undergone a long history of orbital migration. As such the two sub-populations could have accreted the bulk of the their gas in similar locations in the disk, and while hot Jupiters migrated very close to their host star, warm Jupiters did not. In this case we might expect very little chemical difference between the two types of planets.

In what follows we run a similar method as was reported in Paper 1. We compute the astrochemical evolution in the protoplanetary disks that produce each of the warm Jupiters in our model. We then track the abundance of carbon, oxygen, and nitrogen that are available to be accreted into the planetary atmosphere from the disk gas, ice, and refractory sources. We derive the resulting elemental ratios and analyze the connection between these ratios and the physical properties that govern planet formation. We briefly outline our method in \S \ref{sec:method}, report our results in \S \ref{sec:indi}, \ref{sec:results1} and \ref{sec:results2} and discuss the implication on understanding Jupiter-analogs in \S \ref{sec:discussion}. We conclude on this study in \S \ref{sec:conclusion}.

\section{Method: combining astrochemistry and planet formation}\label{sec:method}

As discussed in Paper 1, the main feature of our work is the combination of evolving astrochemical models of protoplanetary disks with a planetesimal accretion model. In this way, we can prescribe the chemical properties (abundances of carbon, oxygen, and nitrogen) in the gas, ice, and refractory components of the protoplanetary disk at the same time and place as the growing proto-planet. The population synthesis model that produced our population of planets is described in \cite{APC2020}. The chemical kinetic code that predicts the gas and ice composition of the disk is based on the work of \cite{Fogel11} and \cite{Cleeves14}, but has been modified for our purposes and described in Paper 1. The chemical model that describes the chemical composition of the refractory component (dust and planetesimals) was introduced in \cite{Crid19b} and includes the possibility of carbon erosion in the inner disk. We outline some of the important concepts here, for more details see \cite{Crid19c} and \cite{APC2020}.

\subsection{ Planet growth and migration }

As a planet grows it evolves through the mass-semi-major axis diagram in Figure \ref{fig:mass_a} through a combination of solid accretion followed by gas accretion (increasing vertically in Figure \ref{fig:mass_a}) and through planetary migration (decreasing horizontally in Figure \ref{fig:mass_a}). \cite{APC2020} uses the planetesimal accretion paradigm \citep{Pollack1996,Ikoma2000,KI02,IL04a,Alibert2005} to build the initial planetary core.The rate of growth is dictated by the surface density of planetesimals which we take as being equal to the dust surface density at any given time. Our dust surface density evolves according to the semi-analytic model of \cite{B12}, primarily through radial drift that quickly empties the outer disk of dust\footnote{In principle the dust surface density also evolves due to the production of planetesimals \citep[for example see][]{Voelkel2020}, however our current implementation is limited as it does not allow such a connection. In practice such a connection will lead to less efficient planetesimal formation and slower initial core growth. Overall this change will not drastically change the main conclusions of the paper. }

Planetesimal formation dictates that the core growth rate is \citep{Pollack1996}:\begin{align}
    \frac{dM_{\rm plnt}}{dt} &= \frac{dM_c}{dt} = \frac{M_c}{\tau_{c,acc}} \nonumber\\
	\simeq & \frac{M_c}{1.2\times 10^5} \left(\frac{\Sigma_{\rm dust}}{10 {\rm gcm}^{-2}}\right)\left(\frac{a}{1 {\rm AU}}\right)^{-1/2} \left(\frac{M_c}{M_\oplus}\right)^{-1/3}\left(\frac{M_s}{M_\odot}\right)^{1/6} \nonumber\\
		\times & \left[\left(\frac{\Sigma_{\rm gas}}{2.4\times 10^3 {\rm gcm}^{-2}}\right)^{-1/5} \left(\frac{a}{1 {\rm AU}}\right)^{1/20}\left(\frac{m}{10^{18} {\rm g}}\right)^{1/15}\right]^{-2} {\rm g~yr}^{-1},
\label{eq:growth}
\end{align}
for a planet core of mass $M_c$ currently orbiting at $a$ around a star of mass $M_s$ accreting planetesimals of (assumed constant) mass $m$. The solid surface density $\Sigma_{\rm dust}$ is determined from the \cite{B12} model, while the gas surface density $\Sigma_{\rm gas}$ is determined by a semi-analytic model based on \cite{Cham09}\footnote{But see \cite{AP18} for the full details of the disk model}.

Once the planet is sufficiently large, it clears the majority of its `feeding zone' of planetesimals and core growth is drastically slowed \citep{IL04a}\footnote{Practically speaking, we increase $\tau_{c,acc}$ by two orders of magnitude in this stage.}. Particularly since the planet migrates through the disk it can continue to accrete planetesimals into its proto-atmosphere, delivering any carbon and oxygen contained within the planetesimal (see below). Due to the reduced planetesimal accretion rate the core begins to cool - which enables a stage of gas accretion to begin \citep{Ikoma2000}. Gas accretion begins at a very slow rate, limited by the Kelvin-Helmholtz timescale \citep{IL04a} such that the mass of the planet evolves as:\begin{align}
    \frac{dM_{\rm plnt}}{dt} = \frac{dM_{\rm gas}}{dt} + \frac{dM_c}{dt},
\end{align}
where $dM_{\rm gas}/dt = M_{\rm plnt}/t_{\rm KH}$, and $dM_c/dt$ proceeds at the aforementioned reduced rate. The Kelvin-Helmholtz time scales with the total mass of the planet \citep{Ikoma2000,AP18}:\begin{align}
    t_{\rm KH} = 10^7 {\rm yr} \left(\frac{M_{\rm plnt}}{M_\oplus}\right)^{-2}.
\end{align}

In the population synthesis model of \cite{APC2020} gas accretion is assumed to halt when the planet reaches some final mass. This final mass is proportional to the gap opening mass with a proportional constant that is generated from a log-normal distribution as part of the population synthesis model. While the general problem of late stage gas accretion remains unsolved, our approach captures the essential points of more complex physical models of the end state of gas accretion \citep[see for example][]{Dangelo2010,Cridland2018}. 

The population synthesis model of \cite{APC2020} stocastically selects a set of the initial disk mass, disk lifetime, and metallicity to initialize the radial distribution of the gas and dust surface densities, the gas temperature, and control the evolution of the disk's mass accretion rate. The initial disk (gas) mass and disk lifetime are selected from a log-normal distribution with an average of 0.1 M$_\odot$ and 3 Myr respectively. Their distribution have a 1$\sigma$ range of 0.073-0.137 M$_\odot$ and 1.8-5 Myr respectively. The disk metallicity ([Fe/H])\footnote{We are using the typical notation where [Fe/H] $\equiv$ $\log_{10}({\rm Fe/H}) - \log_{10}({\rm Fe/H})_\odot$, such that [Fe/H]$_\odot = 0$} is selected from a normal distribution with an average of -0.02 (marginally sub-solar) and a 1$\sigma$ range of -0.22-0.18. The disk metallicity sets the initial gas-to-dust ratio, using the expression:\begin{align}
    f_{\rm gtd} = f_{\rm gtd,0}10^{\rm [Fe/H]},
\end{align}
where $f_{\rm gtd,0}=0.01$ is the typical interstellar medium (ISM) gas-to-dust ratio, such that the radial distribution of dust mass is:\begin{align}
    \Sigma_{\rm dust}(r,t=0) = f_{\rm gtd}\Sigma_{\rm gas}(r,t=0),
\end{align}
where $\Sigma_{\rm gas}$ is derived from the disk model of \cite{Cham09}. Changes in the initial dust surface density impact the rate of the initial core growth through the availability of core-building material at a given ratios.

In Paper 1 we derived the total mass evolution for the set of generated disks and compared them to recent observational surveys of young stellar systems. We found that the population of disks used by \cite{APC2020} reproduced the high-mass end of the observed population of protoplanetary disks, and generally agreed better with the population of Class 0/I objects of \cite{Tychoniec2018}. In this way our generated disks can be thought of as beginning as marginally Class I objects \citep[similar to HL Tau,][]{ALMA15} when we start planet formation, although we ignore the impact of any remaining envelope.

As previously mentioned, growing planets migrate to smaller orbital radii through interactions with the protoplanetary disk gas \citep{LP86,W91}. Planet migration is an ever growing topic since it was first pointed out that the typical timescale for Type-I migration (for low mass planets that do not open gaps) is too short compared to the typical planetesimal accretion timescale to explain the known population of exoplanets \citep{Ward1997}. A way to remedy this discrepancy is to either slow planetary migration, or speed up planetary accretion. The former solution, typically called `planet trapping' posits that discontinuities in the gas density, temperature, or dust opacity can lead to a change in the strength of the torques responsible for migration \citep{Masset06}. This change can slow the migration rate, stop it completely, or even reverse its direction \citep{Masset06,HP10,HP11,McNally2018,McNally2020}. 

In our planet formation model we use the three planet traps outlined in \cite{HP11} - the water ice line, the dead zone edge, and the heat transition - to dictate where the growing planet must be, up until the point where it opens a gap in the disk (discussed below). The particular trap which houses a given planet dictates where that planets begins its core formation. The typical hierarchy is the water ice line being the most inward trap, the dead zone next, and the heat transition beginning the farthest from the host star. The farther from the host star a planets begins the less material is available for core growth, slowing this initial phase of accretion.

Each of the aforementioned planet traps rely on a different transition in the properties of the disk. The water ice line is a transition in the dust opacity located at the sublimation temperature (typically $\sim 170$ K) of water ice. At lower temperatures, water is frozen out onto the dust grains, and their resulting opacity is larger than at larger temperatures where the water is in its vapour phase. \cite{Crid19a} investigated this process in detail and confirmed that such a transition does indeed create a planet trap for the water ice line. Also in \cite{Crid19a}, the dead zone edge (which represents a transition in the disk turbulent $\alpha$) and the heat transition \citep[where the primary heating mechanism changes from viscous to direct irradiation,][]{HP10,Lyra2010} are tested and similarly found to produce planet traps. All three of these traps evolve to smaller radii as the disk loses mass due to its own accretion onto the host star and cools as a result. Hence proto-planets continue to move inward as they grow, but on a much longer timescale (the viscous timescale) than they do in the standard picture of Type I migration. 

Once the planet is sufficiently large it opens a gap in its disk \citep{Crida2009}. At which point Type I migration is suppressed and is replaced by Type II migration \citep{LP86}. During this stage of planetary migration, the planet acts as an intermediary for the angular momentum transport through the disk, and generally migrations inward on the viscous timescale. As we outlined in Paper 1, when the mass of the planet exceeds the gap opening mass then we assume that its radial evolution proceeds on the viscous timescale. 

Apart from adjusting the rate of migration, other methods of saving planets from this `Type-I problem' have been proposed. These include pebble accretion, which posits that the core initial grows through the accretion of cm-sized pebble - a process that can increase the rate of initial core growth by a factor of $\sim 1000$ \citep{Ormel2010,LambJoh2014,Bitsch2015}. Another method involves the direct gravitational collapse of giant planets through an instability at 100s of AU in the protoplanetary disk. While not necessarily faster than core accretion, starting from such large disk radii ensures that there is insufficient time for the growing planet to migrate into the host star.

A final method could simply be that planet formation starts earlier than previously assumed. Recent surveys of protoplanetary disks (also known as Class II objects) have shown that there is insufficient dust currently (by 1-3 Myr) available to produce the core of a Juptier-like planet, let alone multiple planets \citep{Ansdell2016,Manara2018,Tychoniec2018,Tychoniec2020}. Younger Class 0/I objects, however, have been found to contain at least 20$\times$ the dust in Class II objects \citep{Tychoniec2020}. This finding suggests that (at the very least) there is significant planetesimal formation on going in young stellar systems. Planets forming in these systems would likely undergo planet migration, however due to the complex nature of these embedded systems the relevant torques have not yet been characterized.

Here we continue to use our planetesimal accretion and migration prescriptions developed in our past work \citep{HP13,APC16a,Crid16a} and leave the implications of the aforementioned models to future work.

\subsection{ Astrochemistry of the volatiles and refractories }

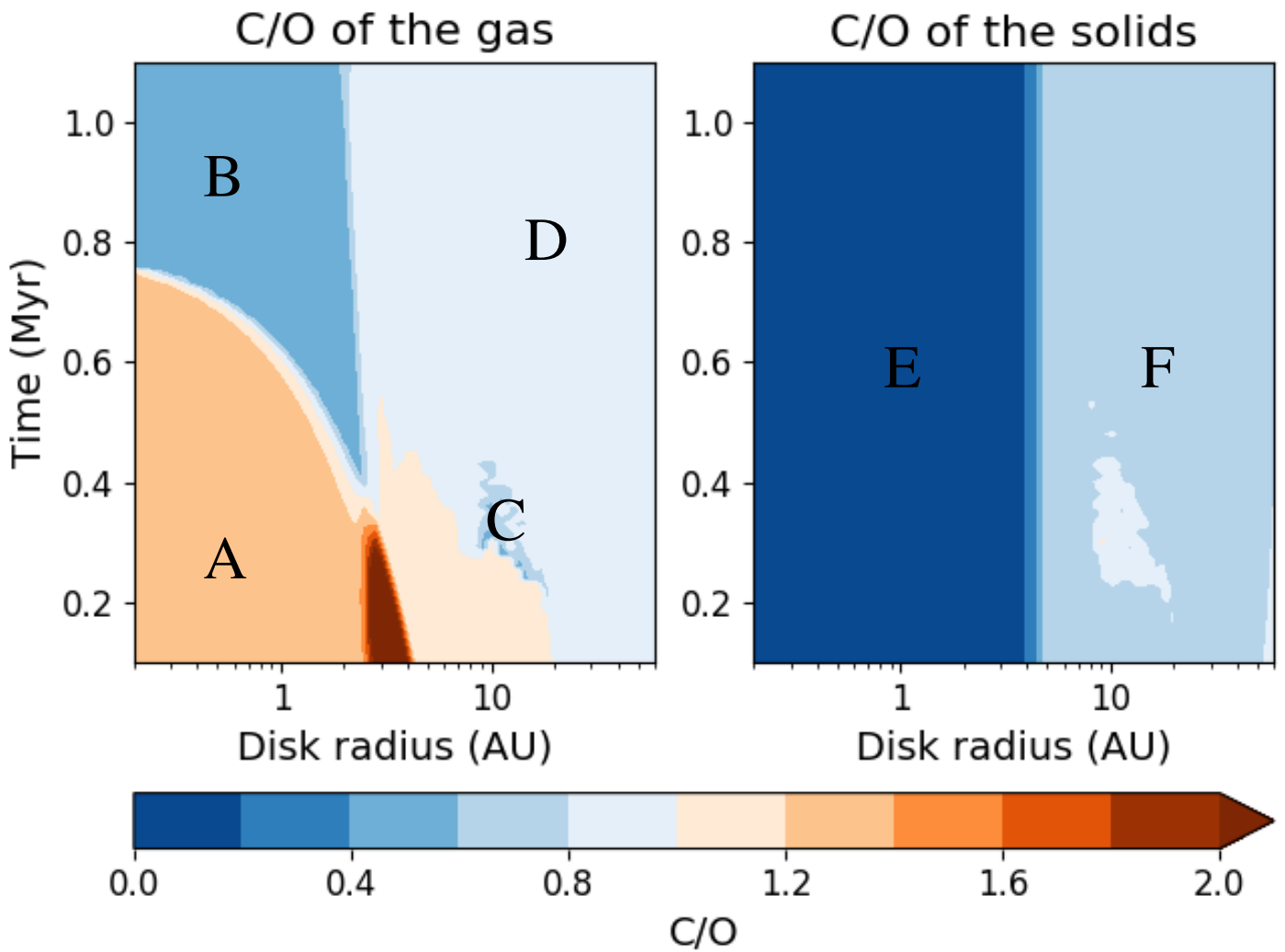
\begin{figure*}
    \centering
    \begin{overpic}[width=\textwidth]{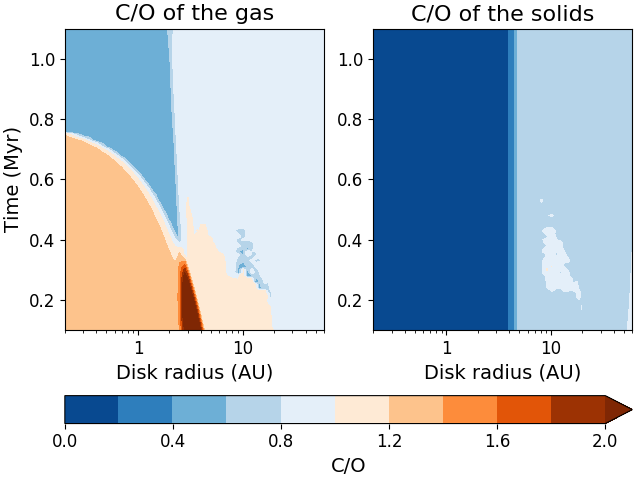}
    \put(15,30){ \Huge A }
    \put(15,60){ \Huge B }
    \put(37,33){ \Huge C }
    \put(40,55){ \Huge D }
    \put(68,45){ \Huge E }
    \put(88,45){ \Huge F }
    \end{overpic}
    \caption{ Evolution and radial distribution of midplane C/O in the gas and solids in one of the disk models used for this work. Specifically we show the chemistry of the disk in the Reset scenario of refractory erosion which governs the high C/O early on in disk life ({\bf A}). We similarly note the region of the disk with high and low water vapour abundances ({\bf B} and {\bf D} respectively), the carbon poor region due to refractory erosion ({\bf E}), and a region where gaseous CO is converted to frozen CO$_2$ ({\bf C}). The carbon-richer region ({\bf F}) exists outward of the carbon erosion front. With this colour scheme, orange denotes carbon-rich regions (C/O $>1$) while blue denotes carbon-poor regions (C/O$<1$). The nominal volatile C/O $=0.4$ while the refractory C/O $=2$.}
    \label{fig:CtoOmap}
\end{figure*}

\subsubsection{Volatiles}

We include an astrochemical model for the evolution of both the volatile and refractory components of the disk carbon, oxygen, and nitrogen. The volatile component of the disk is primary made up of \ce{H2O}, \ce{CO2}, and CO gas and ice - frozen onto dust grains. The disk volatile evolution is computed using the Michigan chemical kinetic code featured in \cite{Fogel11} and \cite{Cleeves14}, and previously used in \cite{Crid16a,Crid17}. It computes the disk chemistry in a 1+1D fashion, assuming vertically isothermal gas and dust, and hydrostatic equilibrium. The chemical evolution is initialized with elemental ratios O/H$_{\rm vol}$ $=2.5\times 10^{-4}$, C/H$_{\rm vol}$ $=1.0\times 10^{-4}$, and N/H$_{\rm vol}$ $=2.45\times 10^{-5}$ assuming an inheritance scenario. Under this scenario the carbon and oxygen begin in their molecular form (largely \ce{CO}, frozen \ce{H2O}) while nitrogen is initialized primarily in atomic N with $\sim 10\%$ molecular \ce{N2}. Under these conditions, the volatile C/O $= 0.4$ and N/O $= 0.098$. The cosmic ray induced ionization rate is $1\times 10^{-17}$/s.

The chemical interaction between the dust grain surface and the gas represents a crucial driver for chemical change. The chemical network underlying the Michigan chemical code includes a limited set of grain surface reactions, primarily focused on the production of molecular hydrogen and water. More complex grain-surface reactions involving carbon-bearing species \citep[as seen in ][]{Walsh15,Eistrup2018,Bosman2018,Krijt2020}, is left out of the chemical model as they typically become relevant at lower temperatures, outside the \ce{CO2} ice line ($\sim 10$ AU). None of our forming warm Jupiters build their atmospheres that far out in the disk. We compute an average dust grain size for our chemical calculation based on the output from the a semi-analytic model of dust evolution \citep{B12}, weighted by the number density of dust grains. For an implementation of this method see \cite{Crid17}, the typical average grain size is $\sim 0.1\mu$m.

The version of the Michigan code that was used in the aforementioned works assumed a passive disk model that remains unchanged over the whole evolution of the chemical system. In Paper 1 we introduced a new version of the code that allowed the disk gas density and temperature to evolve in tandem with the chemistry. This new method introduced new chemical features that did not appear in the passive version of the code (see Paper 1). 

\subsubsection{ Refractories }

A large reservoir of carbon and oxygen also exists in refractory sources - planetesimals and pebbles - in protoplanetary disks \citep{Pon14}. These refractory sources are effectively chemically neutral, and do not contribute to the bulk elemental abundances inferred by (sub)millimeter studies of protoplanetary disks. A possible exception to this trend is carbon, which has shown evidence in our own solar system for an interaction between refractory and volatile sources of the element \citep{Berg15}. Given that the carbon-to-silicon ratio (C/Si) of the ISM is approximately 6, and assuming that the majority of the silicates have the \ce{SiO3} group, then the refractory C/O $=2$. As such, their accretion into the atmospheres of giant planets could have a drastic impact on the resulting C/O \citep[as discussed in][]{Crid19b}. We assume that there is no refractory component for nitrogen.

The Earth is depleted in carbon (relative to silicon) by three orders of magnitude when compared to the carbon-to-silicon ratio of the ISM. Moreover, main-belt asteroids show between one and two orders of magnitude depletion in their C/Si relative to the ISM. This depletion prompted \cite{Berg15} to propose that some chemical process was eroding the carbon off of the dust early in the life of our natal disk - consequently enhancing the gas phase carbon. The chemical processes responsible were investigated by \cite{Lee2010}, \cite{Anderson2017} and \cite{Klarmann2018} but no concrete answer was found. The chemical implication of such a process was investigated by \cite{Wei2019}. They found that the majority of the excess carbon stayed in the gas phase as HCN and hydrocarbons, with only $\sim 1$\% of the carbon condensing back onto the grains in the form of icy long-chain hydrocarbons. 

We include an analytic prescription that describes the distribution and evolution of carbon from the refractory sources into the gas phase. The distribution of the excess gaseous carbon was derived in \cite{Crid19b} and was based on an empirical fit to solar system data by \cite{Mordasini16}. There are two models which describe the distribution of excess carbon: the `reset' and `ongoing' models. As outlined in \cite{Crid19b} these models represent simple but opposing methods for eroding the carbon off the dust grains into the gas. The reset model assumes that during the initial collapse of the molecular cloud a thermal event - similar to a FU Ori outburst - sublimates the dust in the young protoplanetary disk, releasing their contents into the gas phase. As the disk returns to its natural temperature the silicates and iron would recondense into dust, but the carbon would not. This model assumes that all of the erosion necessary to explain today's depleted C/Si of Earth and main-belt asteroids happens at (effectively) $t = 0$. The carbon that would be released due to this process would then advect along with the rest of the gas and dust in the disk into the host star. 

The opposing model, the ongoing model, assumes that there is some ongoing chemical process that is continually eroding carbon off of dust grains in the protoplanetary disk. This process - while not concretely identified - would continually maintain the excess carbon in the disk, as carbon-rich dust grains radially drift into the region of the disk where the erosion can happen \citep[a few AU,][]{Anderson2017}. The main difference between these two models is that the excess carbon vanishes in the reset model after less than 1 Myr \citep[][but also see Figure \ref{fig:CtoOmap}]{Crid19b} while the excess carbon survives the full lifetime of the disk in the ongoing model.

\subsubsection{ General chemical evolution and the carbon erosion front }

In Figure \ref{fig:CtoOmap} we show the radial distribution and evolution of the midplane C/O for both the gas and the solids (ice and refractories) for a single disk model over a span of just over 1 Myr. We note a few points of interest: first we have included the reset model of \cite{Crid19b} which greatly enhances the gaseous carbon content at the expense of carbon from the refractory component. By approximately 0.8 Myr the extra carbon has moved completely into the host star and is no longer available to accrete into any forming proto-planets. Had the ongoing model been included in Figure \ref{fig:CtoOmap}, the carbon rich region A would have extended over all time in much the same was as the carbon poor region E does on the right panel. Note that both the reset and ongoing models result in the carbon poor region E because they both lead to the required depletion in refractory carbon seen currently in the inner solar system. 
The radius where the transition between carbon-richer and carbon poor solids - the carbon erosion front - begins at 5 AU in the fiducial carbon erosion model. The location of the front remains fixed in the ongoing model so that the excess carbon in the gas phase perfectly reflects the depletion of carbon in the solid phase (transition radius between regions E and F). In the reset model, since the excess carbon advects with the bulk gas in the protoplanetary disk the erosion front representing the excess gaseous carbon moves inward. This evolution can be seen in Figure \ref{fig:CtoOmap} as the curved white contour between regions A and B. Later in this work we explore the impact of varying the location of this erosion front.

When the extra carbon due to the reset refractory erosion model advects away from a given disk radius the gas is returned to a lower C/O which is indicative of the initial C/O used in our chemical model (0.4). In region B, inward of the water ice line, the same final C/O can be found as was used as initial conditions. This region slowly shrinks as the disk cools, and the water ice lines moves inward. Outward of the water ice line, in region D, water is primarily in the ice phase, which brings the gas C/O up to a value closer to unity. The carbon and oxygen carrier molecules are dominated by CO in this region, since we only include \ce{CO2} production in the gas phase - which is generally much less efficient than it is in the ice phase (as discussed in Paper 1). We do see a short period of \ce{CO2} production in region C which is produced in the gas before quickly freezing out onto the dust grains at the cost of frozen \ce{H2O} and gaseous CO. As such there is a local decrease in C/O with a subsequent increase of C/O in the solids. This process has already been explored by \cite{Eistrup2016} and was similarly observed in \cite{Crid19c}. However the process lasts only for a few 10$^5$ years before the disk becomes too cold in that region for it to occur efficiently. The slightly more carbon rich region just below C is caused by a small quantity of HCN and long-chain hydrocarbons being produced in the gas phase. We show all of the most abundant gas and ice species in the Appendix figures \ref{fig:chem_test01} and \ref{fig:chem_test02} respectively.

While the left panel of Figure \ref{fig:CtoOmap} shows C/O for the gaseous disk, the right panel shows C/O for the solids. This panel includes C/O for both the ice and refractory sources (dust and planetesimals) but it is dominated by the refractory sources (apart from the small feature mentioned above). This is why it shows much less structure than in left panel - including the decrease in C/O that would accompany the increases seen between regions B and D in the left panel. Instead it mainly shows the transition region inward of the carbon erosion front - the radius where we assume the erosion begin. The carbon refractory erosion model assumes ISM values of carbon (C/H$_{\rm ref} = 2.4\times$ C/H$_{\rm vol}$) outward of the carbon erosion front (assumed to be 5 AU in our fiducial model, more below), while rapidly and smoothly depleting the carbon by a factor of 1000 inside of 5 AU. The functional form of this erosion model was derived empirically by \cite{Mordasini16} and is shown in \cite{Crid19b}.

\subsubsection{ Chemical accounting in planetary atmospheres }
As already discussed, the protoplanetary disk has two sources of carbon, oxygen, and nitrogen for the growing proto-planet. If the gas accretes onto the planet at a rate of $dM_{\rm gas}/dt$ then the total rate of change of any element is simply:\begin{align}
    \frac{dX}{dt} = \frac{1}{\mu m_H}\frac{dM_{\rm gas}}{dt} \times (X/H)_{\rm gas},
\end{align}
where $X/H$ is the abundance of element $X$ relative to hydrogen as computed by our chemical model (volatiles and carbon erosion combined), and $\mu m_H$ is the average weight of a gas particle. In principle micron-sized dust grains will be accreted into the atmosphere along with the gas, since they are well coupled. However along the midplane of the disk (from where we assume the material is accreted) these grains make up a very small fraction of the total mass of the dust (less than 0.1\%). 

Recently, \cite{Cridland2020} explored the impact of vertical accretion on the chemical composition of exoplanetary atmospheres and found that the micron-sized grains can play a role, but only if material is accreted from between one and three gas scale heights. In that case the grains typically brought oxygen-rich ices to the growing planet, generally lowering the atmospheric C/O. For simplicity we ignore the impact of vertical accretion, and hence assume that the micron-sized grains to not contribute to the total mass of the planet nor the chemical structure of its atmosphere.

Conversely we do account for the mass of carbon and oxygen frozen or locked in refractories of the planetesimals that accrete into the proto-atmosphere. For this, we follow the work of \cite{Crid19b} with a simple prescription based on the more detailed calculations of planetesimal survival in planetary atmospheres of \cite{Mordasini15}. We choose an atmospheric mass cutoff of 3 M$_\oplus$ below which an incoming planetesimal survives its trip through the atmosphere, delivering its refractory material directly to the core\footnote{We assume that the core does not contribute to the observed chemical composition of the atmosphere.}. The planetesimal should, however, heat up sufficiently to release any volatiles incorporated in the form of ice. We assume that all volatile (ice) species are released as the planetesimal passes through the atmosphere. 

When the atmospheric mass exceeds 3 M$_\oplus$ \cite{Mordasini15} show that planetesimals (regardless of initial size) will completely evaporate in the atmosphere. For planets that exceed this atmospheric mass we assume that all refractory mass is released and efficiently mixed throughout the atmosphere - thereby impacting the bulk C/O of the planet. We follow \cite{Mordasini16} in assuming that the relative mass fractions of 2:4:3 for carbon (in regions with no carbon erosion), silicates, and irons respectively. As such silicates make up 4/9 of the planetesimal mass in regions with no carbon erosion and 4/7 of the mass in regions where the carbon has eroded.

\section{Results: Individual formation and chemical inheritance } \label{sec:indi}

\begin{figure}
    \centering
    \includegraphics[width=0.5\textwidth]{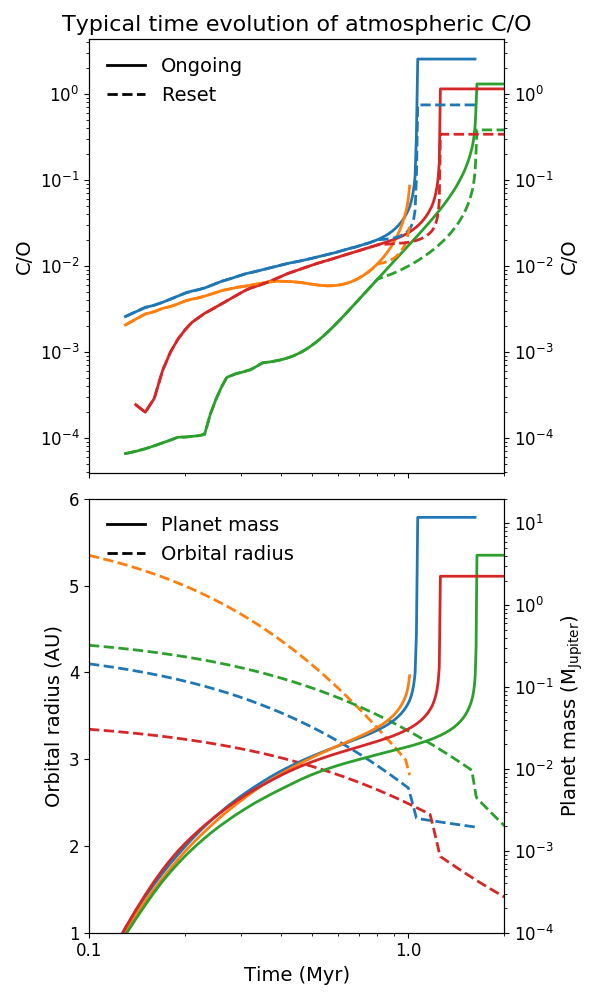}
    \caption{ Typical evolution of atmospheric C/O as a proto-planet grows (top panel). We show the results for each of the example planets for each of the ongoing (solid line) and reset (dashed line) carbon erosion models. In addition we show the mass and orbital radius evolution of the same planets (bottom panel). }
    \label{fig:tracingCtoO}
\end{figure}

To get a sense for the typical evolution of C/O in our growing planets we show, in Figure \ref{fig:tracingCtoO}, the temporal evolution of C/O (top panel) and the orbital radius, and planet mass (bottom panel). The planets begin their evolution as large planetary embryos with M = 0.01 M$_\oplus$. They slowly build up solid mass by accreting planetesimals and a small amount of a gas envelope, building the initial core of M $\sim 10$ M$_\oplus$ in $\sim 0.5$ Myr. At this point the growing planet is slowly accreting gas and any remaining planetesimals (discussed above). By $\sim 1$ Myr the embryos are sufficiently large that they can begin to quickly accrete gas, eventually doing so in an unstable manner. Once the planet has reached its prescribed maximum mass (choosen from a distribution prior to each calculation) its evolution is stopped. 

During the initial build up of the core the `atmospheric' C/O\footnote{That is, the C/O of the envelope that has collected around the proto-planet} is dominated by the release of volatiles frozen onto incoming planetesimals as they pass through the early proto-atmosphere. Planets forming near or inward of the water ice line are underabundant in volatiles, but can contain small amounts of hydrocarbons. These, when combined with the small amount of gas (which is rich in water vapour) that is accreted at this stage leads to the low atmospheric C/O for the first $\sim$Myr. Planets forming outward of the water ice line accrete planetesimals rich in frozen water which drives very low initial C/O. All four of these planets grow in the transition region of the carbon erosion model, meaning that planets forming closer to the host star accrete planetesimals with less carbon than planets forming farther away. Once their proto-atmosphere is sufficiently large, refractories begin to contribute to the atmospheric C/O and hence planets forming farther away (green line) see a steeper increase in C/O than planets closer in (red and blue lines) after $\sim 0.5$ Myr. Because it forms outward of the water ice line, the planet denoted by the orange line sees its C/O slightly reduced prior to the beginning of more rapid gas accretion.

Once gas accretion starts to dominate the mass evolution, the atmospheric C/O begins to evolve towards the local C/O of the protoplanetary disk, this includes the impact of the carbon erosion model. In the case of the reset model (dashed line) the atmospheric C/O evolves towards $\sim 0.4$ inward of the water ice line (green and red lines) and $\sim 0.7$ outward of the water ice line (blue line). Planets accreting in short-lived disks (orange line) are stranded at lower C/O because their gas accretion is halted by the photoevaporating disk. In the case of the ongoing model there is extra carbon in the gas that is accreted by the growing planets, enhancing their atmospheric C/O.

Because of the unstable gas accretion in our formation model, most of the C/O evolution happens over a short period of time - as the bulk of the atmosphere is accreted. As such, giant planets freeze in the chemical composition of the gas at their location in the protoplanetary disk where their unstable gas accretion occurred. Lower mass planets (orange line) do not undergo unstable gas accretion and as such the history of their solid accretion can be more important to their final atmospheric C/O.

\section{Results: warm Jupiters compared to hot Jupiters} \label{sec:results1}

To follow up our study of hot Jupiters in Paper 1, a natural question to ask was whether these synthetic warm Jupiters share any chemical similarities to the hot Jupiters. To that end we follow a similar trajectory as in Paper 1 here and outline some general properties of our population of planets.

\subsection{Orbital and mass distribution}

\begin{figure}
    \centering
    \includegraphics[width=0.5\textwidth]{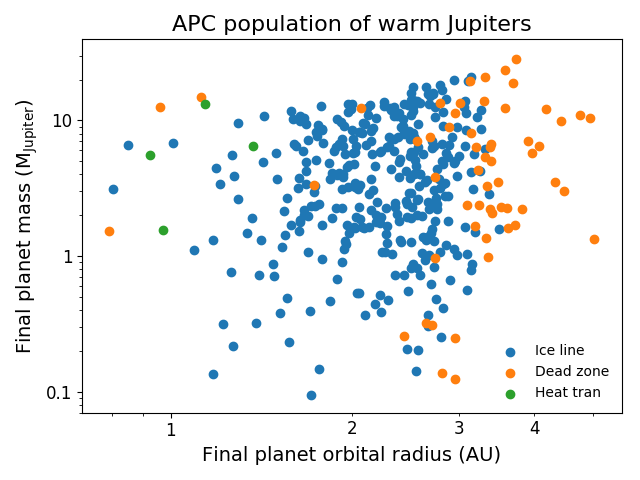}
    \caption{Final mass and orbital range of synthetic planets from the \cite[][APC]{APC2020} population of planets. The colour coding here (and throughout) denotes the planet trap in which the planet initially grew. Generally ice line planets (blue) start closer than the dead zone (orange) and heat transition (green) planets. The population of warm Jupiters exist predominately between 1-4 AU.}
    \label{fig:massorb}
\end{figure}

In Figure \ref{fig:massorb} we show the distribution of the final orbital radius (assuming circular orbits) and mass of the population of warm Jupiters from \cite{APC2020}. These planets orbit between 0.5 - 5 AU, with the vast majority orbiting between 1 - 4 AU. The masses range from a third of Saturn's mass ($\sim 0.1$  M$_{\rm Jupiter}$) up to $\sim 30$ M$_{\rm Jupiter}$. As such, this population of planets extends into the mass range that is typically associated with brown dwarfs stars. In what follows we will not differentiate between brown dwarfs and planets in our analysis, since this distinction is irrelevant for our analysis of bulk elemental abundance ratios.

We colour code each planet by the planetary trap from which it originates. The majority of these planets arise from the water ice line trap (blue). This trap is an optimal location for the generation of planetesimals \citep{Drazkowska2017}, which is depicted in our model by an enhancement in the dust surface density at the water ice line caused by a `traffic jam' effect \citep[see][for a discussion of this effect]{Pinilla2016,Crid16b}. The dead zone edge (orange) generally begins farther outward than the water ice line, and generally leads to planets which end their formation farther from their host star than those from the water ice line. A few exceptions to this trend exist, and these planets generally emerge in disks with longer lifetimes. In our model, longer lived disks also evolve slower, hence the surface density and temperature reduce slower which keeps the dead zone edge at larger radii for longer. Planets trapped at the dead zone edge see lower densities than planets that begin closer to the host star and they can migrate further inward before they begin to accrete large amounts of gas which ends their formation closer to the host star than the average dead zone planet.

For a similar reason, planets originating from the heat transition (heat tran, green) trap tend to be larger and closer-in than the majority of the water ice line planets. Generally speaking, planets forming in the heat transition trap produce super-Earth planets \citep{APC2020}. The planets formed here grew from protoplanetary disks at the low-mass end of our disk mass distribution which caused the initial location of the heat transition to be closer to the host star than would be usual. In our model, the heat transition evolves on the viscous timescale, while the dead zone edge evolves slightly faster \citep{APC16a}. Because of its slower radial evolution, more time passes before the growing planet reaches a higher density environment where its growth can proceed more quickly. As such its final radii are generally farther inward of the bulk of the water ice line planets.

\subsection{Connection to Jupiter}

As already mentioned, our population of warm Jupiters does not include Jupiter and Saturn in their current orbital states. However our population does include a fair number of planets in a range of orbital radii indicative of Jupiter just prior to undergoing a possible Grand Tack \citep{Walsh2011}. As such we consider Jupiter-analogs to be planets with a similar mass as Jupiter, but at an orbital radius closer to their host star - having missed a Grand Tack. Either because its planetary system lacks a companion, or because the mass and/or orbital radii ratios were not tuned to complete a successful Grand Tack.

\subsection{C/O of warm Jupiters and Jupiter-analogs}

\begin{figure}
    \centering
    \includegraphics[width=0.5\textwidth]{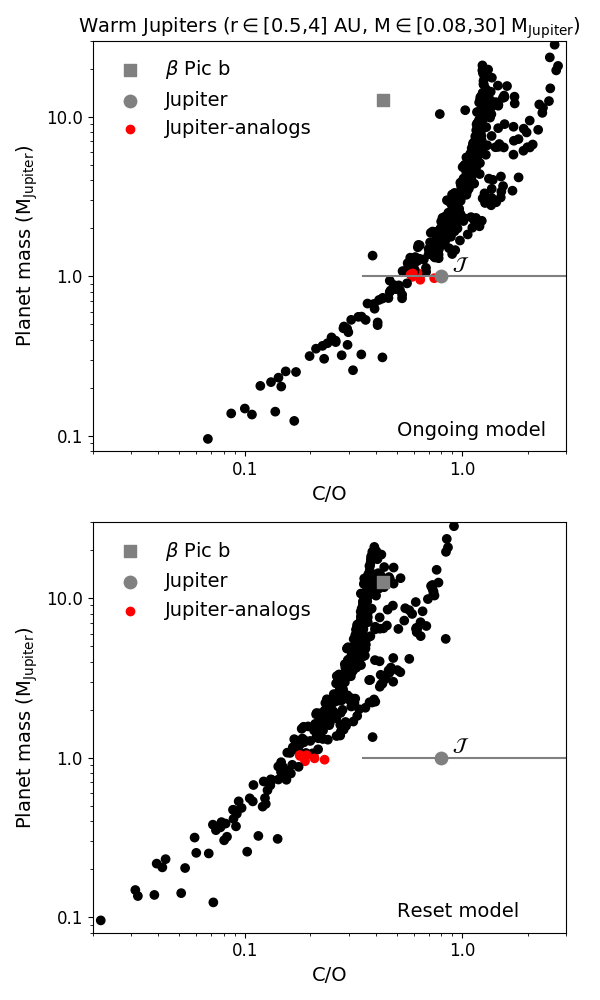}
    \caption{The resulting C/O for our population of planets as a function of mass (black points). We note the Jupiter-analogs from our population in red and observational data in grey. The data point for $\beta$ Pic comes from \citet{GRAVITY2020}, and the data for Jupiter from \citet{Asplund2009} and \citet{Li2020}. The gray line accompanying Jupiter's C/O (noted with $\mathcal{J}$) shows the range of possible values based on the 1$\sigma$ uncertainty of C/H \citep{Asplund2009} and O/H \citep{Li2020}. }
    \label{fig:CtoOzone3}
\end{figure}

In Figure \ref{fig:CtoOzone3} we show the resulting C/O for our population of warm Jupiters, the Jupiter-analogs are highlighted with orange points. The main difference in the resulting atmospheric C/O between the reset and ongoing carbon erosion models is a horizontal shift in C/O for the majority of planets (although not all, explored below). The ongoing model, with its constant production of excess carbon, results in more carbon-rich planets compared to the reset model. There is a small discrepancy in this observation for a few planets that do not migrate inward of the erosion front (at 5 AU) until after they have accreted the majority of their atmosphere. These planets are difficult to see here, but are highlighted and discussed in the following section.

We include an estimated C/O for Jupiter based on the measurements outlined in \cite{Asplund2009} and the recent oxygen measurement of \cite{Li2020}. The error bars on this measurement are computed with the maximum and minimum C/H and O/H provided by the 1$\sigma$ uncertainty in the above papers. Clearly there is a wide range of possible C/O based on these uncertain measurements. Including these uncertainties, Jupiter is most consistent with the Jupiter-analogs in the ongoing carbon erosion model. This suggests both that Jupiter accretes the majority of its gas inward of the carbon erosion front (inward of its current orbit), and that there was an ongoing chemical process responsible for the processing of carbon off of dust grains throughout the life of the solar nebula. This conclusion would hold if only carbon and oxygen are considered. Nitrogen and the noble gasses, however, have recently suggested that Jupiter's initial growth (and at least a fraction of its gas accretion) occurred outward of the \ce{N2} ice line - at tens of AU \citep{Bosman2019,Oberg2019}. As is discussed in more detail below, there is a clear discrepancy between Jupiter's C, N, and O elemental abundances when they are combined in comparison with the presented population of warm Jupiters.

\ignore{
Apart from these aforementioned planets, chemical data for planets of this type is lacking. One candidate planet, WAPS-167e was identified by \cite{WASP167e} using a \textit{Spitzer} transmission detection at 3.6 $\mu$m. Orbiting at a radius of 1.87 AU \citep{WASP167e} it is one of the only planets of its kind to have an IR transmission signal detected. While its mass is not well constrained, its orbital period has been confirmed to very high accuracy. Because of its precise orbital period, further transmission studies are possible and would provide a unique look into the chemical properties of this otherwise common population of exoplanets.

 We note this discrepancy here, however, to connect our population to $\beta$ Pic b.
 }

In addition to Jupiter we have included the recent measurement of $\beta$ Pic b made by \cite{GRAVITY2020}. $\beta$ Pic b is a $\sim 12$ M$_{\rm Jupiter}$ planet orbiting near 10 AU around its host star. Its C/O was determined from a pair of retrieval models based on interferometric observations of its atmosphere by \cite{GRAVITY2020}. Given its orbital radius, it lies on the outer edge of what we defined as warm Jupiters. Indeed, planetesimal accretion in general, and our formation models in particular struggle to make large planets at these larger radii - and a formation scheme like pebble accretion \citep{Ormel2010,Johansen2007,Bitsch2015} may be better suited to explain their existence. Regardless we find that its measured C/O is consistent with planets that formed in the reset carbon refractory erosion model. This fact argues for its formation to have occurred in a region of the disk outward of both the water ice line and the refractory carbon erosion front. For more discussion regarding the carbon erosion front see Section \ref{sec:erosion}. Outward of the carbon erosion front the gas is less carbon rich than it could be inward of the front, but the solids are more carbon rich (recall Figure \ref{fig:CtoOmap}, right panel). \ignore{While the initial core growth of $\beta$ Pic b is likely more consistent with pebble accretion, its consistency with our chemical models does suggest that gas accretion and late stage pollution of the atmosphere by planetesimal accretion is probably the same everywhere in the disk - and only the initial core growth differs between formation models.}

\subsection{The C/O main sequence and mass-metallicity relation}

As in Paper 1, we wish to understand the cause of the structure we see in Figure \ref{fig:CtoOzone3}. One major conclusion from Paper 1 was the C/O main sequence, which shows a tight inverse correlation between the fraction of the total mass made up of solids with the atmospheric C/O. 

\begin{figure}
    \centering
    \includegraphics[width=0.5\textwidth]{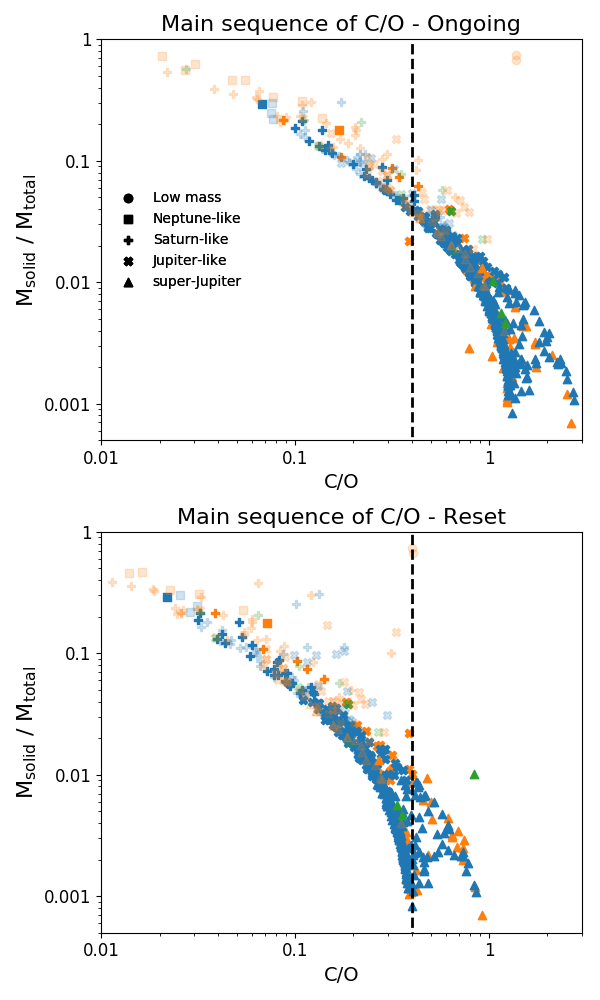}
    \caption{The C/O main sequence for both the ongoing and reset carbon erosion models. The vertical dashed lines show C/O $=0.4$ that initialized the chemical model for the volatile component of the disk. The data from Paper 1 are also included as faded points to show that indeed the population of warm Jupiters follows the general trend of the main sequence from Paper 1. At the very high C/O end the population appears to drop away from the trend found at lower C/O. The planets in this part of the figure are very large mass and are dominated by gas accretion. The colour of each point denotes the trap from which the planet originated: ice line (blue), dead zone (orange), and heat transition (green).}
    \label{fig:mainseq}
\end{figure}

In Figure \ref{fig:mainseq} we show the same main sequence as presented in Paper 1, with the data from Paper 1 included as faded points. As was done in Paper 1, we differentiate between different planet masses; the mass bins are: low mass (M $<$ 10 M$_\oplus$), Neptune-like (10 M$_\oplus$ $<$ M $<$ 40 M$_\oplus$), Saturn-like (40 M$_\oplus$ $<$ M $<$ 200 M$_\oplus$), Jupiter-like (200 M$_\oplus$ $<$ M $<$ 790 M$_\oplus$), and super-Jupiter (790 M$_\oplus$ $<$ M). We find that the population of warm Jupiters tend to follow the main sequence up to high C/O where it then falls away from the trend. The high C/O end is dominated by the most massive planets (super-Jupiters) with masses even higher than were obtained in Paper 1. For these most massive planets, their atmospheric chemistry is determined almost entirely by gas accretion (with solid accretion contributing less than 1\%). As such, in the reset model (bottom panel) one group of the massive planets tend towards C/O of the disk volatiles used in the chemistry calculation (vertical dashed line) while the other tends to higher C/O. This trend is linked to where the planets accreted their gas - inward of the water ice line the C/O tends to the disk C/O while outward of the water ice line the C/O tends to unity. These groupings are not discrete, however, and there are planets which seem to exist between the two extremes. We explore these groupings in more detail in section \ref{sec:breakdown}. The same structure can be seen for the ongoing model (top panel) but it is shifted to higher C/O caused by the excess carbon that remains in the gas for the whole lifetime of the disk.

\begin{figure}
    \centering
    \includegraphics[width=0.5\textwidth]{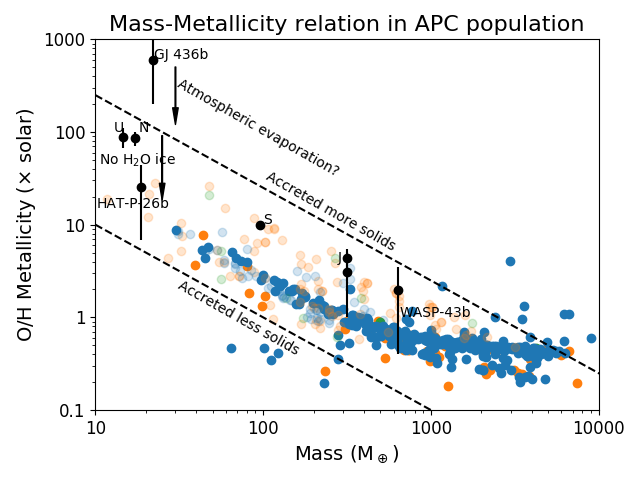}
    \caption{The mass-metallicity relation for the \cite[][APC]{APC2020} population of warm Jupiters, compared to the hot Jupiter population from Paper 1 (faded points). The solar system giants (inferred by methane abundance and taken from \citealt{Kre14}), and for WASP-43 b \citep{Kre14}, GJ 436 b \citep{Morley2017}, and HAT-P-26 b \citep{MacDonald2019} (inferred from their water abundance) are also shown. In addition we include the recent O/H measurement for Jupiter by \cite{Li2020} to show that, within uncertainty, the relation is independent of using C/H or O/H to determine the metallicity. We include O/H for our synthetic population and find that they follow the relation up to a mass of a few Jupiter masses where they appear to flatten out. The colour of each point denotes the trap from which the planet originated: ice line (blue), dead zone (orange), and heat transition (green). This is a recreation of Figure 12a. from Paper 1. }
    \label{fig:massmetal}
\end{figure}

In Figure \ref{fig:massmetal} we present the mass-metallicity relation for the population of warm Jupiters and compare them directly to the population of hot Jupiters from Paper 1. We find that (as in Paper 1), the mass-metallicity relation directly follows from the main sequence - that is, the atmosphere metallicity falls with increasing planet mass. There is, in addition, a similar turn off of the main trend at the higher mass end, with massive planets (of at least a few Jupiter masses) tending towards a metallicity of between 0.4-0.5 $\times$ solar. The low mass end of the population align very closely to a region of O/H - mass parameter space that we attribute to planets having accreted their gas outward of the water ice line (in an oxygen-poor region of the gas disk). In general most of the lower mass warm Jupiters sit lower than planets of similar mass in the hot Jupiter population suggesting that this is a general trend. This is a reflection of the fact that warm Saturn and Neptune planets largely accreted their gas outside of the water ice line. We explore further causes of scatter seen here in a following section.

\section{Results: exploring different elemental ratios}\label{sec:results2}

While the C/O ratio gives an important view of planet formation, it is not the only elemental ratio that can shed light on the problem. After carbon and oxygen, nitrogen is the next most abundant in the solar system. As discussed in \cite{Bosman2019}, nitrogen chemistry is generally very simple - since the majority of the element remains in its molecular form \ce{N2} at gas temperatures lower than $\sim 500$ K while at higher temperatures \ce{NH3} becomes dominant (see Figure \ref{fig:chem_test01}). Apart from this transition, there is the small build up of HCN that was previously mentioned, which can hold on the order of 1\% of the total nitrogen. Recall that we attribute this HCN enhancement to the carbon rich region just below region C in Figure \ref{fig:CtoOmap}. 

\begin{figure}
    \centering
    \includegraphics[width=0.5\textwidth]{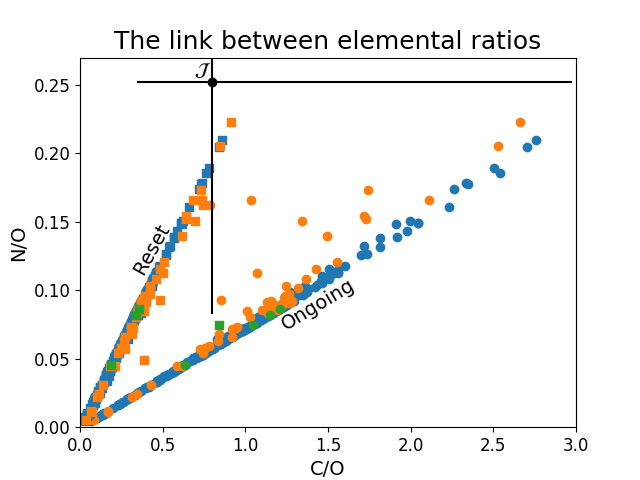}
    \caption{The cross reference of the C/O with N/O for the population of warm Jupiters. We show here both the reset (squares) and ongoing (circles) models which generally lie on a pair of straight lines. The slope of these lines represent the elemental ratio N/C which must be generally constant for most planets. The black point and error bars denote Jupiter's C/O and N/O ratios along with estimates for possible ranges. Due to the uncertainty in observed elemental abundances Jupiter is consistent with planets from the reset model but not with the ongoing model. The colour of each point denotes the trap from which the planet originated: ice line (blue), dead zone (orange), and heat transition (green).}
    \label{fig:CtoONtoO}
\end{figure}

In Figure \ref{fig:CtoONtoO} we explore the role that nitrogen can play in understanding the physics of planet formation. Here we plot the nitrogen-to-oxygen ratio (N/O) against the C/O ratio for our population of warm Jupiters and for both the reset (square) and ongoing (circle) models. Note that since we separately run the reset and ongoing models for each planet formed in our model, each planet has two points on this figure - one for each carbon erosion model. We immediately see that the majority of planets fall on two straight lines - one for each of the carbon erosion models. Given that the slope of these lines is the nitrogen-to-carbon ratio (N/C) we can say that for the planets in our model, the N/C ratio is effectively constant. There are a number of planets, however, that do not conform to this rule and they can be grouped into carbon-rich planets in the reset model and carbon-poor planets in the ongoing model.

We additionally place Jupiter on Figure \ref{fig:CtoONtoO} using the elemental abundances reported in \cite{Asplund2009} and the new oxygen abundance from \cite{Li2020}. While it seemed to be consistent with our population of Jupiter-analogs accreting from the ongoing model in Figure \ref{fig:CtoOzone3}, Jupiter does not fit well into their C/O vs. N/O parameter space. Within uncertainty, Jupiter's C/O and N/O ratios are consistent with the set of planets forming under the reset carbon erosion model. The inconsistency between its fit in the mass-C/O parameter space (more consistent with the ongoing model) and its fit here (with the reset model) suggests that its formation is inconsistent with the formation of warm-Jupiters through planeteimsal formation presented here.

This inconsistency provides further evidence that Jupiter formed farther outward in the disk than is achieved by our model, furthering a growing belief that was proposed in \cite{Oberg2019} and \cite{Bosman2019}. They place Jupiter's initial formation location outward of the \ce{N2} ice line at tens of AU away from the Sun, likely formed through the accretion of icy pebbles. Here we note that while the planets in our population typically most of their carbon from gas sources, \cite{Bosman2019} propose that Jupiter's carbon content is largely accreted from frozen CO accompanying the accreting pebbles.

\begin{figure}
    \centering
    \includegraphics[width=0.5\textwidth]{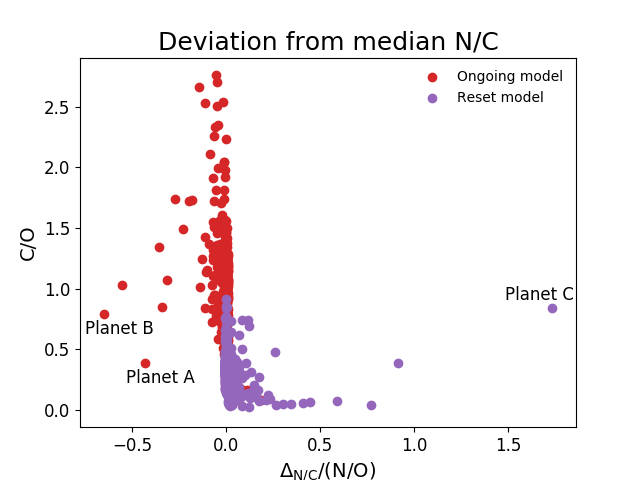}
    \caption{The deviation of N/O from the straight lines shown in Figure \ref{fig:CtoONtoO} which represents a constant N/C across all planets in the population. For further analysis we label three planets that show the wide deviation from the median N/C of the population.}
    \label{fig:NtoCmeddiff}
\end{figure}

Returning to our population of planets, to quantify the distance off the general trend in Figure \ref{fig:CtoONtoO} of a constant N/C, we compute the deviation away from the general trend for all planets in both erosion models relative to N/O. To do this we first compute a median N/C (N/C$_{\rm median}$) for each of the ongoing and reset model results. We then assume that the connection between N/O and C/O can be explained simply by a linear function with slope equal N/C$_{\rm median}$. Deviation from this general trend would have the form: \begin{align}
    \Delta_{\rm N/C} = {\rm N/O} - {\rm N/C}_{\rm median} \cdot {\rm C/O},
    \label{eq:01}
\end{align}
where N/O and C/O are the elemental ratios computed by our model. The absolute value of $\Delta_{\rm N/C}$ and its sign shows how far from the line with slope $N/C_{\rm median}$ and in what direction.

We show the result of this calculation in Figure \ref{fig:NtoCmeddiff}. The majority of the points lie within 0.01 of $\Delta_{\rm N/C} = 0$, meaning that their computed elemental abundances are consistent with the average planet in our population. There are a few planets which show larger deviations in Figures \ref{fig:CtoONtoO} and \ref{fig:NtoCmeddiff}, and we select three of these planets to further investigate. Planet A  lies to the left of the average planet in the ongoing model. Its elemental abundances do not appear to depend on the erosion model in which it forms (its points overlap in Figure \ref{fig:CtoONtoO}). Planet B similarly sits far to the left of the average planet forming in the ongoing model. Finally Planet C is a planet that lies far to the right of an average planet from the reset model. Similar to the previous two planets, it shows similar C/O and N/O ratios when it is formed in both the reset and ongoing carbon erosion models.

\begin{figure*}
\begin{subfigure}{0.333\textwidth}
\centering
    \includegraphics[width=\textwidth]{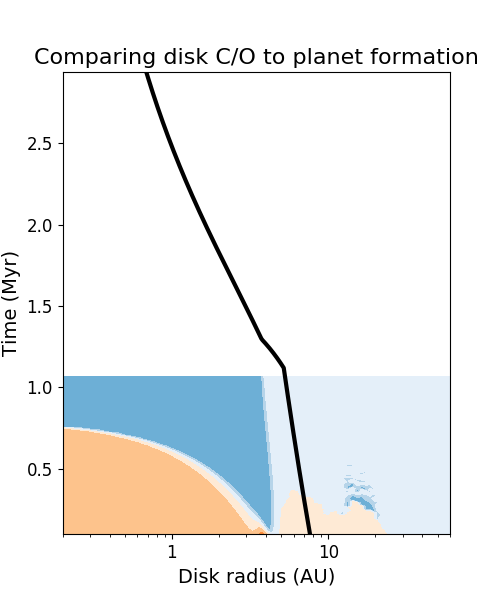}
    \caption{Planet A from Figure \ref{fig:NtoCmeddiff}.}
        \label{fig:PlanetA}
\end{subfigure} 
\begin{subfigure}{0.333\textwidth}
 \centering
    \includegraphics[width=\textwidth]{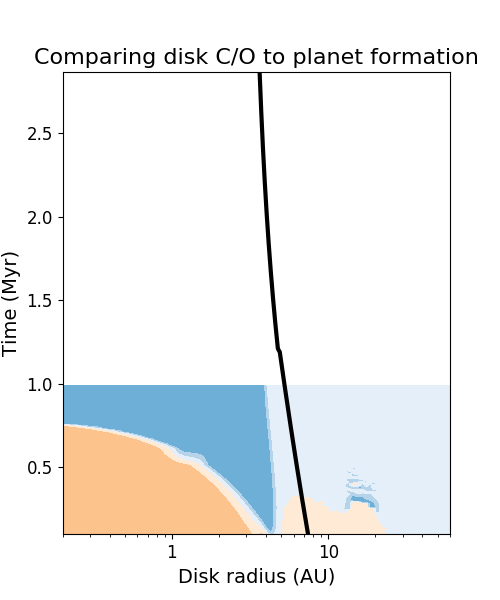}
    \caption{Planet B from Figure \ref{fig:NtoCmeddiff}.}
    \label{fig:PlanetB}
\end{subfigure} 
\begin{subfigure}{0.333\textwidth}
     \includegraphics[width=\textwidth]{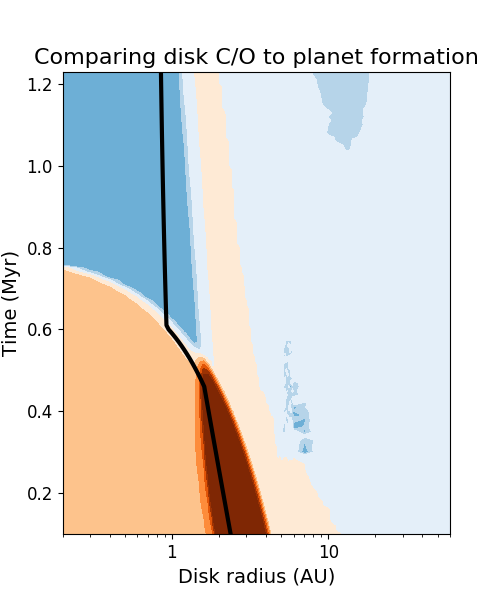}
    \caption{Planet C from Figure \ref{fig:NtoCmeddiff}.}
    \label{fig:PlanetC}
\end{subfigure}
\caption{Comparison between the planets A, B, and C's location (black line) and the underlying chemical properties of the gas (coloured contours). Note we only show C/O up to the point where gas accretion is shut off in our formation model, since the disk no longer impacts the atmosphere once accretion is stopped. The contours are the same as in Figure \ref{fig:CtoOmap}. Note the change in time axis in the third panel.}
\label{fig:Planets}
\end{figure*}

\ignore{
\begin{figure}
    \centering
    \includegraphics[width=0.5\textwidth]{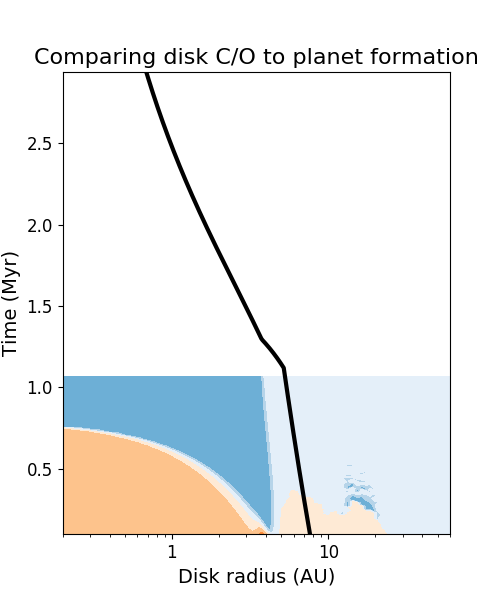}
     \caption{Comparison between the Planet A's location (black line) and the underlying chemical properties of the gas (coloured contours). Note we only show C/O up to the point where gas accretion is shut off in our formation model, since the disk no longer impacts the atmosphere once accretion is stopped. The contours are the same as in Figure \ref{fig:CtoOmap}.}
        \label{fig:PlanetA}
\end{figure}

\begin{figure}[h!]
    \begin{minipage}{0.5\textwidth}
    \centering
    \includegraphics[width=\textwidth]{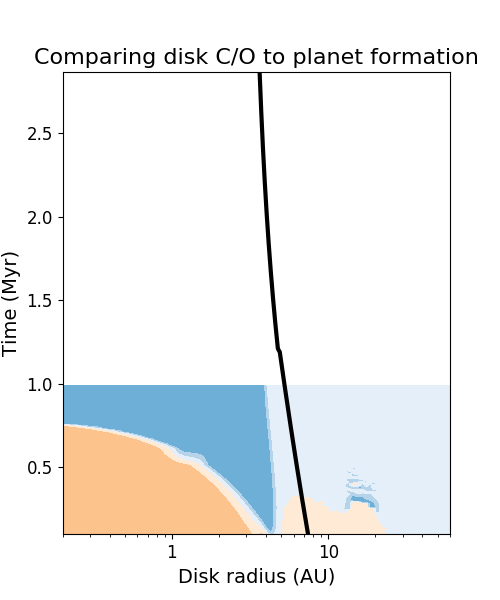}
    \caption{Same as in Figure \ref{fig:PlanetA} but for Planet B.}
    \label{fig:PlanetB}
    \end{minipage}
    \begin{minipage}{0.5\textwidth}
    \centering
    \includegraphics[width=\textwidth]{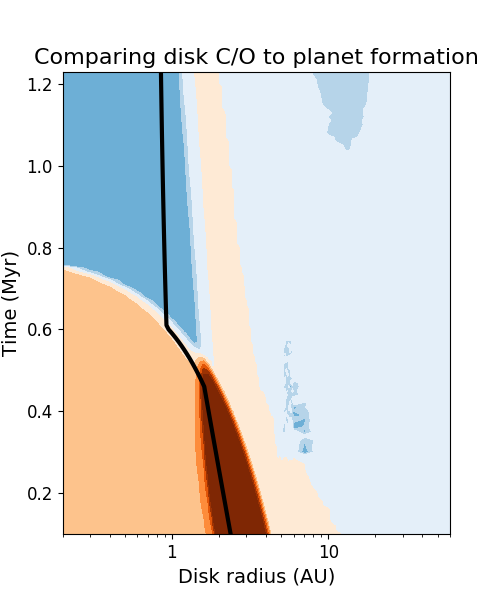}
    \caption{Same as in Figure \ref{fig:PlanetA} but for Planet C.}
    \label{fig:PlanetC}
    \end{minipage}
\end{figure}
}

In Figures \ref{fig:PlanetA} - \ref{fig:PlanetC} we compare the radial evolution for each of these planets with the underlying gas C/O. To reflect the time frame that is most important for setting the chemical composition of each of their atmospheres we only show contours up to the time where the planet's growth is truncated in our planet formation model. Figure \ref{fig:PlanetA} and \ref{fig:PlanetB} show very similar pictures for Planets A and B. They each began growing farther out than the carbon erosion front (at 5 AU) and never crossed the front until after gas accretion has been terminated. As such they both have fed on predominately oxygen-poorer (0.8 $<$ C/O $<1$) gas and slightly oxygen-richer (0.6 $<$ C/O $<0.8$) solids. 

While the two planets accrete nearly the same amount of solids in total ($\sim 9$ M$_\oplus$) the main difference between them is that Planet B ends up accreting roughly an order of magnitude more gas than Planet A. This difference comes from the randomly generated maximum mass parameter from the population synthesis model, with Planet B being aloud to accrete for longer than Planet A. As such the planet accreted more gas which lead to the chemistry in the atmosphere of Planet B being more dependent on gas accretion than Planet A. A combined measurement of C/O and N/O can help to understand the formation history of a planet and/or whether its natal disk underwent a refractory carbon erosion-like process.

In Figure \ref{fig:PlanetC} we compare the gas chemistry and orbital migration history for Planet C. We can see two important features in both the chemical composition of Planet C's disk as well as the migration of the planet. The marginally carbon-rich region of the disk (below region C in Figure \ref{fig:CtoOmap}) extends for much longer in time in this disk than in the disk shown in Figure \ref{fig:CtoOmap}. The water ice line is also closer to the host star than in Figure \ref{fig:CtoOmap} which is a property of colder, less massive disks. Planet C begins its formation inward of the carbon erosion front, and coincidental evolves inward at the same rate (and at the same disk radius) as the erosion front in the reset model between $\sim 0.5$ - $0.65$ Myr.

Because of Planet C's orbital coincidence with the carbon erosion front, it spends a large portion of its formation accreting carbon-rich gas - even in the reset model - as such it ends it formation with a very large C/O for its N/O. In addition, we find that Planet C's C/O is nearly independent of the carbon erosion model, because it spends a sufficiently long time accreting carbon-rich gas in the reset model. The difference in C/O between the two carbon erosion models for Planet C is only about 20\% - much smaller than for a typical planet in our population.

\section{Discussion: What sets C/O in warm-Jupiter atmospheres?} \label{sec:discussion}

So far, we have reported on our findings for the C/O and N/O for either the entire population of warm Jupiters, or on an individual level. Generally we have found (as was the case in Paper 1) that the fraction of mass that is accreted as solids into the atmosphere tends to heavily constrain the chemical properties that result. There are some exceptions, however, where the formation history - particularly the migration history - also has a noticeable impact on the resulting chemical properties of the atmosphere. In this section we divide the population of planets by their initial disk conditions and study the resulting C/O in the context of the environment in which they form.

\subsection{Break down of C/O by protoplanetary disk properties}\label{sec:breakdown}

\begin{figure*}
    \centering
    \includegraphics[width=\textwidth]{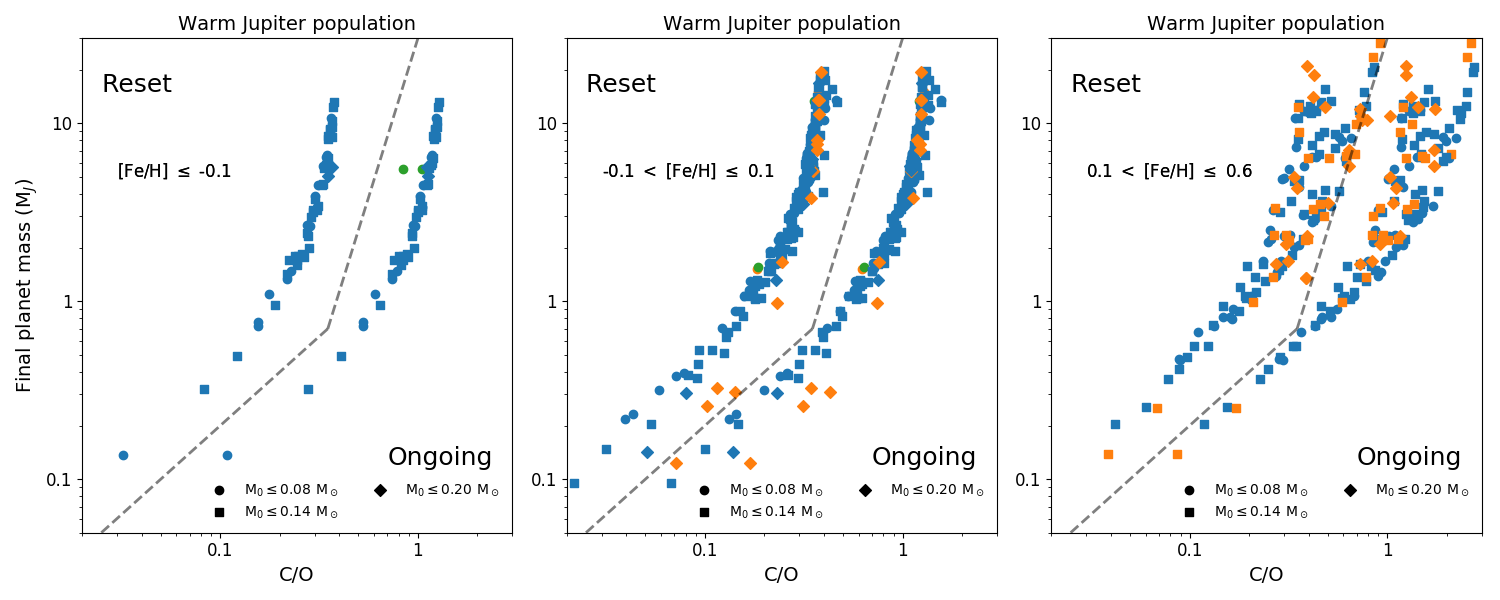}
    \caption{C/O for synthetic planets in both of the ongoing and reset carbon erosion models. The data for each erosion model are separated on each panel by a dashed line. Each panels separates the planets by the protoplanetary disk metallicity. The left panel represents low metallicity, the middle panel is solar-like metallicity ($-0.1<$[Fe/H]$\leq 0.1$), and the right panel represents high metallicity ($0.1<$[Fe/H]$\leq 0.6$). The different point shapes represent different initial disk masses.  The colour of each point denotes the trap from which the planet originated: ice line (blue), dead zone (orange), and heat transition (green). Generally there is a shift in C/O between planets growing in the reset and ongoing carbon erosion models. However there are some exceptions: for example the heat transition planet (green point) in the first panel which has nearly the same C/O in both the reset and ongoing model (this is Planet C from above).}
        \label{fig:CtoOvsmany0}
\end{figure*}

In Figures \ref{fig:CtoOvsmany0} we split the C/O results of both the ongoing and reset models (respectively) into groups of disk metallicity, initial disk mass, and occupying planet trap. These groups are denoted by separate panels, marker shape, and colour respectively. The left panel of the figure denotes the metal-poor systems ([Fe/H] $\leq$ -0.1), the middle panel denotes solar-like metallicities ($-0.1<$ [Fe/H] $\leq 0.1$), and the right panel denotes metal-rich systems ($0.1<$ [Fe/H] $\leq 0.6$). In all figures we place dashed lines to differentiate between the atmospheric results for planets forming in the reset and ongoing models.

In Figure \ref{fig:CtoOvsmany0} we bin the ongoing and reset model C/O data as discussed above. In the metal-poor and solar-like panels the planets lie tightly correlated over a wide range of mass and C/O ratios. The tight correlation implies that planets forming in disks with metal-poor and solar-like metallicities are the systems that most consistently produce planets that agree with the main sequence of mass-C/O ratio introduced in Paper 1.

Generally speaking, planets trapped at the dead zone edge and heat transition traps require the largest and lightest disk masses respectively to form warm Jupiters in our formation model. This is due to the timescale related to the initial core build up, which needs to be sufficiently fast to build giant planets within the lifetime of the disk. The low mass disks tend to be cooler which moves the heat transition inward to smaller radii and relatively higher densities than would be present in higher mass disks. Conversely the dead zone trap best builds planets in disks with initially higher masses. The dead zone edge is computed semi-analytically in \cite{APC2020} and is less sensitive to the initial disk mass as is the heat transition, hence higher disk masses lead to higher densities at the trap and faster core growth. For lower mass disks at these metallicities, dead zone trapped planets lead more often to hot Jupiters. 

The metal-rich panel of planets is the first to show significant deviations from the general trends of the two other panels. As already discussed in relation to Figure \ref{fig:mainseq}, a second group of warm Jupiters have higher C/O than would be predicted from their planet mass or fraction of mass accreted as solids. This group is more evenly spread when binning by metallicity than was suggested in Figure \ref{fig:mainseq}, showing that the metal-rich systems lead to higher chemical diversity than the lower metallicity systems. Given their high C/O the most likely scenario for these planets are that they accreted the majority of their gas outward of the water ice line. This is most easily done in the metal-rich disks because there is a higher density of solids (by construction) when the metallicity is higher, which reduces the timescale related to the initial core growth. In these systems core formation can occur efficiently farther away from the host star than in the lower metallicity systems. This generates a wider variety of chemical properties as the planets sample chemically different regions of the disk. This variety causes some of the scatter seen in Figure \ref{fig:massmetal} because their O/H metallicity is constrained by the accretion of generally oxygen-poorer gas outward of the water ice line.

In Figure \ref{fig:CtoOvsmany0} we also show the reset model. Generally the difference between the two figures is a shift of all points to less carbon-rich atmospheres in the reset model, apart from a few particular planets (three of which were discussed above). In particular the heat transition planet in the top panel stands out as having the highest C/O in that group (this was Planet C from above), and the dead zone planet in the bottom panel that lies the lowest along the right dashed line (this was Planet A from above). Otherwise the structure of the metal-poor and solar-like groups of planets are grouped by planet mass and C/O in a similar way as in the ongoing model.

In the metal-rich panel we see a slight change in the structure of the distribution of planets. The dead zone trapped planets tend to be more carbon-rich than the planets coming from the water ice line trap in the reset model than was seen in the ongoing model. As previously argued, the planets forming at the dead zone edge tend to start their evolution farther from the host star than the water ice line planets. As such they accrete carbon-richer gas than is found inward of the water ice line. In the reset model the excess carbon is lost to the host star after less than 0.8 Myr and planets forming farther outward in the disk are more sensitive to the volatile chemistry than in the ongoing model. 

Overall we find that for solar-like and metal-poor disks, there is a reasonably tight correlation between the planet mass and the C/O. This implies that the main-sequence derived from the population of hot Jupiters extends easily to the warm-Jupiters with much of the scatter at the high planet mass end being produced by metal-rich systems. These metal-rich systems were not found in Paper 1 because high-metallicity disks tend to build warm Jupiters over hot Jupiters in our formation model. At the low planet mass end of the distribution, there appears to be a small deviation in C/O caused by differences in initial disk mass. This is particularly apparent in the solar-like metallicity systems, where we see that planets generated in low mass disks tended to be less carbon-rich than planets of the same planet mass forming from high mass disks. 

\subsection{Impact of shifting the carbon erosion front} \label{sec:erosion}

For the majority of the paper, and the entirety of Paper 1, we have assumed that the process of carbon erosion began at 5 AU, with the excess carbon smoothly increasing to 1 AU, inward of which we maintained a constant carbon excess. This assumption, however, is largely based on current observations of the refractory component of carbon in the Earth mantle, asteroids, comets, and Jupiter's current orbital radius \citep{Berg15,Mordasini16}. Jupiter very likely migrated to its current location, either from smaller radii during a Grand Tack, or from larger radii. Indeed Jupiter's migration is \textit{required} to explain the current population of Trojan asteroids \citep{Pirani2019}.

Since the chemical process that drives refractory erosion is still an open question, and Jupiter very likely (must have) migrated during its formation then it is completely reasonable to vary the radial location of the carbon erosion front. For simplicity, we anchor the inner region of the function that describes the carbon erosion, such that the excess carbon is always constant inward of 1 AU. We then vary the carbon erosion front from between 3 - 7 AU. This radius range ensures that the shifting erosion front remains relevant for the presented population of planets.

The location of the erosion front impacts the chemical properties of the disk in two ways. First it changes the partitioning of the excess carbon between the gas and refractories through the location where carbon is removed from the grains. Moving the front inward implies that there is generally less carbon available in the gas for accretion, and a wider range of radii where the refractories remain carbon-rich. In the reset model, the second effect is the timing at which the excess carbon is lost to the host star. We keep the advection speed the same throughout this work, so moving the front inward shortens the time it takes the excess carbon to accrete into the host star. The opposite is true if we move the erosion radius farther away from the host star.

\begin{figure}
    \centering
    \includegraphics[width=0.5\textwidth]{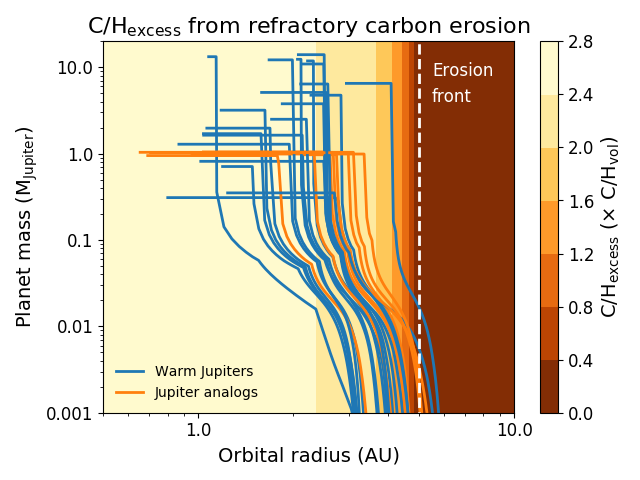}
    \caption{A few selected planetary tracks compared to the location of the erosion front and the resulting distribution of excess carbon (in contours).}
    \label{fig:erosion}
\end{figure}

For illustrative purposes, in Figure \ref{fig:erosion} we show a comparison between a few of our planet tracks (which describe the planet's evolution through the mass-semi-major axis diagram), the carbon erosion front, and the resulting excess carbon generated from the refractories (contours). We show the distribution of the excess carbon in the ongoing model for an erosion front of 5 AU (fiducial model). Clearly, if the front was shifted inward then the excess carbon available to some planets will be reduced, while if it is moved outward then a higher carbon excess is available to some of the growing planets earlier in their formation. 

\begin{figure}
    \centering
    \includegraphics[width=0.5\textwidth]{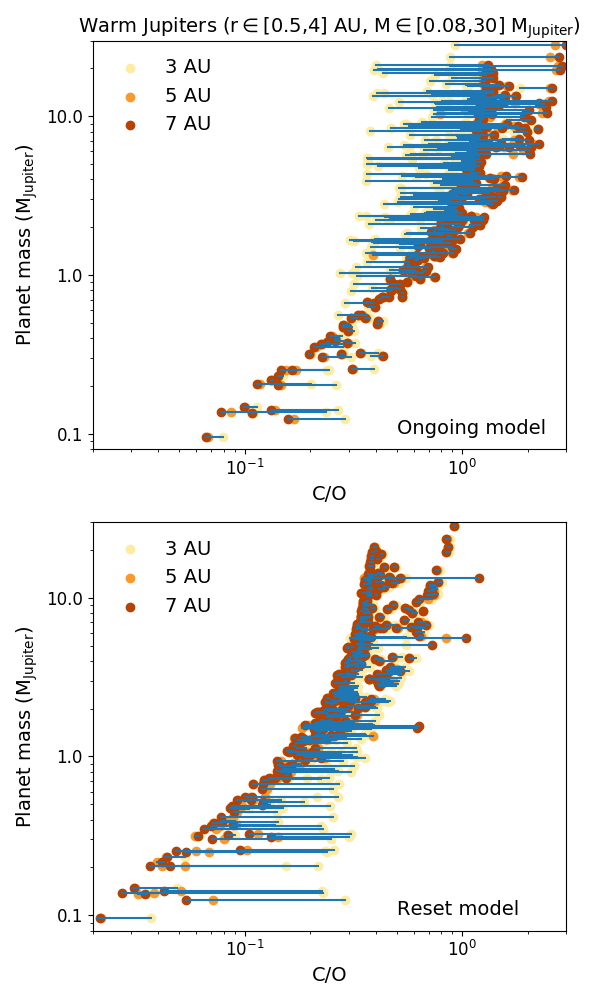}
    \caption{Same as Figure \ref{fig:CtoOzone3}, but including the effect of shifting the erosion to between 3 AU and 7 AU.}
    \label{fig:shifted}
\end{figure}

In Figure \ref{fig:shifted} we show the impact of shifting the carbon erosion front between 3 - 7 AU for both the ongoing (top panel) and reset (bottom panel) models. The planets forming in the ongoing model show two distinct shifts in relation to the change in the carbon erosion front. Lower mass planets (${\rm M} \lesssim 0.55 {\rm M}_{\rm Jupiter}$) shift to higher C/O when the carbon erosion front is shifted to smaller radii, while the opposite is true for higher mass planets. This difference highlights how higher mass planets depend more on gas accretion for setting their chemical composition while lower mass planets are more dependent on solid accretion. In a sense, a better classification of `ice giants' are planets with M $\lesssim 0.55 {\rm M}_{\rm Jupiter} \sim 175 {\rm M}_\oplus$ and gas giants as planet with ${\rm M} > 175 {\rm M}_\oplus$. Although we admit that such a classification would be confusing as it would change our solar system to contain one gas giant and three ice giants, nevertheless such a classification would better capture the physics of and chemistry of planet formation. 

In the bottom panel of Figure \ref{fig:shifted} we see that the majority of the planets are shifted to more carbon-rich atmospheres when the carbon erosion front is moved inward. This is because there is more carbon in total available for planetary accretion in the reset model, since fewer solids are chemically processed and less carbon is lost to the host star. The lower mass planets see the highest shift since their chemical composition is most dependent on the refractory source of carbon. There are a few planets where the opposite trend is observed, with higher C/O for a carbon erosion front farther away from the host star. Like Planet C from earlier, this is caused by a coincidence between the excess carbon in the gas and the period of rapid gas accretion during the planets' formation.

\section{Conclusion} \label{sec:conclusion}

Here we have presented a population of warm Jupiters, derived from a full planet population synthesis model starting from planetesimal accretion. We computed the chemical evolution of the disk volatiles in each of the disks used for the population synthesis calculation to predict the carbon, oxygen, and nitrogen abundances in the gas and ice. When combined with a model for the refractory chemistry - particularly the carbon - we compute the evolution of the total carbon, oxygen, and nitrogen content of each protoplanetary disk. These calculations are combined with the planet tracks derived from the formation model to predict the resulting C/O and N/O in the planetary atmospheres.

\noindent We generally find that:
\begin{itemize}
    \item Like the hot Jupiters in Paper 1 \citep{Crid19c} there is a reasonably tight correlation between the C/O in the atmosphere and the planetary mass.
    \item The spread in the aforementioned correlation is linked to planets forming in metal-rich disks, which are capable of producing more chemically varied atmospheres due to a more rapid initial core build up.
    \item The main sequence of the C/O ratio vs. fraction of solid mass accreted into the atmosphere reported in Paper 1 is upheld for the warm Jupiters. High mass planets tend to curve away from the general trend, moving asymptotically towards the disk volatile C/O in the case of the reset model, and to C/O $\sim 1$ for the ongoing model.
    \item There is an arm of higher C/O caused by the metal-rich disks seen in the main sequence that asymptote to a higher C/O in both carbon erosion models.
    \item Including N/O into our analysis weakens the viability of planetesimal formation in the inner disk as the formation mechanism of Jupiter. This result agrees with the purely chemical analysis of \cite{Oberg2019} and \cite{Bosman2019} which places Jupiter's formation origin outward of tens of AU.
    \item Combining C/O and N/O additionally allows us to identify planets that exclusively accrete their atmosphere outside of the refractory carbon erosion front - a rare situation for this population of planets. 
    \item Shifting the carbon erosion front shows the importance of solid accretion in determining the chemical structure of planetary atmospheres particularly for planets with mass $\lesssim 175$ M$_\oplus$. 
\end{itemize}

The observability of these types of planets will continue to be a challenge as they lie in a range of orbital period that make their chemical characterization difficult by both transit spectroscopy as well as direct imaging. There is a single exoplanet, WASP-167e, that has had its orbital period (of 1071 days) characterized by both Kepler and \textit{Spitzer} with enough accuracy to justify an attempt for transit spectroscopy with JWST \citep{WASP167e}. As we enter into the next generation of Extremely Large Telescopes, and with improvements to coronography that are ongoing, it is possible that a direct emission spectrum of warm Jupiters could be taken. This could unlock a whole new range of planets to study chemically.

Along with the observational challenges, there is more to be done on the modelling side of planet formation and astrochemistry. This work has made strides to include a simple model that describes the chemical properties of the refractories in the protoplanetary disks. The physics that describe how this material finds its way into the atmosphere of the planet, and where the carbon and oxygen are deposited in the atmosphere still remain complicated problems that are beyond the scope of this paper. The strict atmospheric mass cutoff that governs the delivery of refractory material into the atmosphere is simply implemented in our model (as described in Paper 1). Self-consistently computing the arrival and destruction of planetesimals into the atmosphere of a giant planet could deliver refractory material deep enough into the atmosphere that it is unable to impact the observable C/O ratios. Such a complication is currently beyond the scope of this work.

Furthermore the protoplanetary disk models that are used in our planet formation models are smooth - representing a much simpler picture than is being seen in current high resolution surveys of young star forming regions. These surveys are also pushing the start time for planet formation - or at least the formation of the first planetesimals - farther back into the Class 0 or Class I young stellar systems \citep{Tychoniec2018}. It is perfectly possible that by the time a Class II disk (classically called a protoplanetary disk, as it was believed to be the natal system for planets) emerges from the envelope of the proto-star that planets have already almost fully formed. Indeed the marginally Class I/II system HL Tau already shows wide shallow gaps that are often attributed to the presence of at least one large planet \citep{ALMA15,Tamayo15}. If it is indeed true that the majority of the initial core growth and atmosphere accretion (important for determining the chemistry of the atmosphere) occurs in the Class 0/I phase, then we will need adjust our disk models to incorporate the properties of these systems.

With all that being said, it is safe to say that there is still much that can be learned about the physical processes governing planet formation from models like the ones presented here. And over the next few decades, as we begin to chemically characterize the atmospheres of planets as easy as it is now to find them; we should see another surge in our understanding of planet formation. A surge that could rival the one we witnessed in the early years of the \textit{Kepler} mission, possibly driven by JWST or through the upcoming European Space Agency's ARIEL mission.

\begin{acknowledgements}

 Astrochemistry in Leiden is supported by the European Union A-ERC grant 291141 CHEMPLAN, by the Netherlands Research School for Astronomy (NOVA). The work made use of the Shared Hierarchical Academic Research Computating Network (SHARCNET: www.sharcnet.ca) and Compute/Calcul Canada. A.J.C acknowledges additional support by the European Union ERC grant H2020 ExoplanetBio supervised by Ignas Snellen. R.E.P. is supported by an NSERC Discovery Grant. M.A. acknowledges funding from NSERC through the PGS-D Alexander Graham Bell scholarship.

\end{acknowledgements}


\bibliographystyle{aa} 
\bibliography{mybib.bib} 

\begin{thebibliography}{89}
\expandafter\ifx\csname natexlab\endcsname\relax\def\natexlab#1{#1}\fi

\bibitem[{{Alessi} \& {Pudritz}(2018)}]{AP18}
{Alessi}, M. \& {Pudritz}, R.~E. 2018, \mnras, 478, 2599

\bibitem[{{Alessi} {et~al.}(2017){Alessi}, {Pudritz}, \& {Cridland}}]{APC16a}
{Alessi}, M., {Pudritz}, R.~E., \& {Cridland}, A.~J. 2017, \mnras, 464, 428

\bibitem[{{Alessi} {et~al.}(2020){Alessi}, {Pudritz}, \& {Cridland}}]{APC2020}
{Alessi}, M., {Pudritz}, R.~E., \& {Cridland}, A.~J. 2020, \mnras, 493, 1013

\bibitem[{{Alibert} {et~al.}(2005){Alibert}, {Mousis}, {Mordasini}, \&
  {Benz}}]{Alibert2005}
{Alibert}, Y., {Mousis}, O., {Mordasini}, C., \& {Benz}, W. 2005, \apjl, 626,
  L57

\bibitem[{{ALMA Partnership} {et~al.}(2015){ALMA Partnership}, {Brogan},
  {P{\'e}rez}, {Hunter}, {Dent}, {Hales}, {Hills}, {Corder}, {Fomalont},
  {Vlahakis}, {Asaki}, {Barkats}, {Hirota}, {Hodge}, {Impellizzeri}, {Kneissl},
  {Liuzzo}, {Lucas}, {Marcelino}, {Matsushita}, {Nakanishi}, {Phillips},
  {Richards}, {Toledo}, {Aladro}, {Broguiere}, {Cortes}, {Cortes}, {Espada},
  {Galarza}, {Garcia-Appadoo}, {Guzman-Ramirez}, {Humphreys}, {Jung}, {Kameno},
  {Laing}, {Leon}, {Marconi}, {Mignano}, {Nikolic}, {Nyman}, {Radiszcz},
  {Remijan}, {Rod{\'o}n}, {Sawada}, {Takahashi}, {Tilanus}, {Vila Vilaro},
  {Watson}, {Wiklind}, {Akiyama}, {Chapillon}, {de Gregorio-Monsalvo}, {Di
  Francesco}, {Gueth}, {Kawamura}, {Lee}, {Nguyen Luong}, {Mangum}, {Pietu},
  {Sanhueza}, {Saigo}, {Takakuwa}, {Ubach}, {van Kempen}, {Wootten},
  {Castro-Carrizo}, {Francke}, {Gallardo}, {Garcia}, {Gonzalez}, {Hill},
  {Kaminski}, {Kurono}, {Liu}, {Lopez}, {Morales}, {Plarre}, {Schieven},
  {Testi}, {Videla}, {Villard}, {Andreani}, {Hibbard}, \& {Tatematsu}}]{ALMA15}
{ALMA Partnership}, {Brogan}, C.~L., {P{\'e}rez}, L.~M., {et~al.} 2015, \apjl,
  808, L3

\bibitem[{{Anderson} {et~al.}(2017){Anderson}, {Bergin}, {Blake}, {Ciesla},
  {Visser}, \& {Lee}}]{Anderson2017}
{Anderson}, D.~E., {Bergin}, E.~A., {Blake}, G.~A., {et~al.} 2017, \apj, 845,
  13

\bibitem[{{Ansdell} {et~al.}(2016){Ansdell}, {Williams}, {van der Marel},
  {Carpenter}, {Guidi}, {Hogerheijde}, {Mathews}, {Manara}, {Miotello},
  {Natta}, {Oliveira}, {Tazzari}, {Testi}, {van Dishoeck}, \& {van
  Terwisga}}]{Ansdell2016}
{Ansdell}, M., {Williams}, J.~P., {van der Marel}, N., {et~al.} 2016, \apj,
  828, 46

\bibitem[{{Asplund} {et~al.}(2009){Asplund}, {Grevesse}, {Sauval}, \&
  {Scott}}]{Asplund2009}
{Asplund}, M., {Grevesse}, N., {Sauval}, A.~J., \& {Scott}, P. 2009, \araa, 47,
  481

\bibitem[{{Aumann} {et~al.}(1969){Aumann}, {Gillespie}, \& {Low}}]{Aumann1969}
{Aumann}, H.~H., {Gillespie}, C.~M., J., \& {Low}, F.~J. 1969, \apjl, 157, L69

\bibitem[{{Bergin} {et~al.}(2015){Bergin}, {Blake}, {Ciesla}, {Hirschmann}, \&
  {Li}}]{Berg15}
{Bergin}, E.~A., {Blake}, G.~A., {Ciesla}, F., {Hirschmann}, M.~M., \& {Li}, J.
  2015, Proceedings of the National Academy of Science, 112, 8965

\bibitem[{{Birnstiel} {et~al.}(2012){Birnstiel}, {Klahr}, \& {Ercolano}}]{B12}
{Birnstiel}, T., {Klahr}, H., \& {Ercolano}, B. 2012, \aap, 539, A148

\bibitem[{{Bitsch} {et~al.}(2015){Bitsch}, {Lambrechts}, \&
  {Johansen}}]{Bitsch2015}
{Bitsch}, B., {Lambrechts}, M., \& {Johansen}, A. 2015, \aap, 582, A112

\bibitem[{{Bosman} {et~al.}(2019){Bosman}, {Cridland}, \&
  {Miguel}}]{Bosman2019}
{Bosman}, A.~D., {Cridland}, A.~J., \& {Miguel}, Y. 2019, \aap, 632, L11

\bibitem[{{Bosman} {et~al.}(2018){Bosman}, {Walsh}, \& {van
  Dishoeck}}]{Bosman2018}
{Bosman}, A.~D., {Walsh}, C., \& {van Dishoeck}, E.~F. 2018, \aap, 618, A182

\bibitem[{{Brewer} \& {Fischer}(2016)}]{Brewer2016b}
{Brewer}, J.~M. \& {Fischer}, D.~A. 2016, \apj, 831, 20

\bibitem[{{Brogi} {et~al.}(2014){Brogi}, {de Kok}, {Birkby}, {Schwarz}, \&
  {Snellen}}]{Brogi2014}
{Brogi}, M., {de Kok}, R.~J., {Birkby}, J.~L., {Schwarz}, H., \& {Snellen},
  I.~A.~G. 2014, \aap, 565, A124

\bibitem[{{Chambers}(2018)}]{Chambers2018}
{Chambers}, J. 2018, \apj, 865, 30

\bibitem[{{Chambers}(2009)}]{Cham09}
{Chambers}, J.~E. 2009, \apj, 705, 1206

\bibitem[{{Chametla} {et~al.}(2020){Chametla}, {D'Angelo}, {Reyes-Ruiz}, \&
  {S{\'a}nchez-Salcedo}}]{Chametla2020}
{Chametla}, R.~O., {D'Angelo}, G., {Reyes-Ruiz}, M., \& {S{\'a}nchez-Salcedo},
  F.~J. 2020, \mnras, 492, 6007

\bibitem[{{Cleeves} {et~al.}(2014){Cleeves}, {Bergin}, \& {Adams}}]{Cleeves14}
{Cleeves}, L.~I., {Bergin}, E.~A., \& {Adams}, F.~C. 2014, \apj, 794, 123

\bibitem[{{Crida}(2009)}]{Crida2009}
{Crida}, A. 2009, \apj, 698, 606

\bibitem[{{Cridland}(2018)}]{Cridland2018}
{Cridland}, A.~J. 2018, \aap, 619, A165

\bibitem[{{Cridland} {et~al.}(2020){Cridland}, {Bosman}, \& {van
  Dishoeck}}]{Cridland2020}
{Cridland}, A.~J., {Bosman}, A.~D., \& {van Dishoeck}, E.~F. 2020, \aap, 635,
  A68

\bibitem[{{Cridland} {et~al.}(2019{\natexlab{a}}){Cridland}, {Eistrup}, \& {van
  Dishoeck}}]{Crid19b}
{Cridland}, A.~J., {Eistrup}, C., \& {van Dishoeck}, E.~F. 2019{\natexlab{a}},
  \aap, 627, A127

\bibitem[{{Cridland} {et~al.}(2016){Cridland}, {Pudritz}, \&
  {Alessi}}]{Crid16a}
{Cridland}, A.~J., {Pudritz}, R.~E., \& {Alessi}, M. 2016, \mnras, 461, 3274

\bibitem[{{Cridland} {et~al.}(2019{\natexlab{b}}){Cridland}, {Pudritz}, \&
  {Alessi}}]{Crid19a}
{Cridland}, A.~J., {Pudritz}, R.~E., \& {Alessi}, M. 2019{\natexlab{b}},
  \mnras, 484, 345

\bibitem[{{Cridland} {et~al.}(2017{\natexlab{a}}){Cridland}, {Pudritz}, \&
  {Birnstiel}}]{Crid16b}
{Cridland}, A.~J., {Pudritz}, R.~E., \& {Birnstiel}, T. 2017{\natexlab{a}},
  \mnras, 465, 3865

\bibitem[{{Cridland} {et~al.}(2017{\natexlab{b}}){Cridland}, {Pudritz},
  {Birnstiel}, {Cleeves}, \& {Bergin}}]{Crid17}
{Cridland}, A.~J., {Pudritz}, R.~E., {Birnstiel}, T., {Cleeves}, L.~I., \&
  {Bergin}, E.~A. 2017{\natexlab{b}}, \mnras, 469, 3910

\bibitem[{{Cridland} {et~al.}(2019{\natexlab{c}}){Cridland}, {van Dishoeck},
  {Alessi}, \& {Pudritz}}]{Crid19c}
{Cridland}, A.~J., {van Dishoeck}, E.~F., {Alessi}, M., \& {Pudritz}, R.~E.
  2019{\natexlab{c}}, \aap, 632, A63

\bibitem[{{Dalba} \& {Tamburo}(2019)}]{WASP167e}
{Dalba}, P.~A. \& {Tamburo}, P. 2019, \apjl, 873, L17

\bibitem[{{D'Angelo} {et~al.}(2010){D'Angelo}, {Durisen}, \&
  {Lissauer}}]{Dangelo2010}
{D'Angelo}, G., {Durisen}, R.~H., \& {Lissauer}, J.~J. 2010, {Giant Planet
  Formation}, ed. S.~{Seager}, 319--346

\bibitem[{{Dodson-Robinson} {et~al.}(2009){Dodson-Robinson}, {Veras}, {Ford},
  \& {Beichman}}]{DR2009}
{Dodson-Robinson}, S.~E., {Veras}, D., {Ford}, E.~B., \& {Beichman}, C.~A.
  2009, \apj, 707, 79

\bibitem[{{Dr{\c a}{\.z}kowska} \& {Alibert}(2017)}]{Drazkowska2017}
{Dr{\c a}{\.z}kowska}, J. \& {Alibert}, Y. 2017, \aap, 608, A92

\bibitem[{{Eistrup} {et~al.}(2016){Eistrup}, {Walsh}, \& {van
  Dishoeck}}]{Eistrup2016}
{Eistrup}, C., {Walsh}, C., \& {van Dishoeck}, E.~F. 2016, \aap, 595, A83

\bibitem[{{Eistrup} {et~al.}(2018){Eistrup}, {Walsh}, \& {van
  Dishoeck}}]{Eistrup2018}
{Eistrup}, C., {Walsh}, C., \& {van Dishoeck}, E.~F. 2018, \aap, 613, A14

\bibitem[{{Fogel} {et~al.}(2011){Fogel}, {Bethell}, {Bergin}, {Calvet}, \&
  {Semenov}}]{Fogel11}
{Fogel}, J.~K.~J., {Bethell}, T.~J., {Bergin}, E.~A., {Calvet}, N., \&
  {Semenov}, D. 2011, \apj, 726, 29

\bibitem[{{Gandhi} \& {Madhusudhan}(2018)}]{Gandhi2018}
{Gandhi}, S. \& {Madhusudhan}, N. 2018, \mnras, 474, 271

\bibitem[{{Gravity Collaboration} {et~al.}(2019){Gravity Collaboration},
  {Lacour}, {Nowak}, {Wang}, {Pfuhl}, {Eisenhauer}, {Abuter}, {Amorim},
  {Anugu}, {Benisty}, {Berger}, {Beust}, {Blind}, {Bonnefoy}, {Bonnet},
  {Bourget}, {Brandner}, {Buron}, {Collin}, {Charnay}, {Chapron}, {Cl{\'e}net},
  {Coud{\'e} Du Foresto}, {de Zeeuw}, {Deen}, {Dembet}, {Dexter}, {Duvert},
  {Eckart}, {F{\"o}rster Schreiber}, {F{\'e}dou}, {Garcia}, {Garcia Lopez},
  {Gao}, {Gendron}, {Genzel}, {Gillessen}, {Gordo}, {Greenbaum}, {Habibi},
  {Haubois}, {Hau{\ss}mann}, {Henning}, {Hippler}, {Horrobin}, {Hubert},
  {Jimenez Rosales}, {Jocou}, {Kendrew}, {Kervella}, {Kolb}, {Lagrange},
  {Lapeyr{\`e}re}, {Le Bouquin}, {L{\'e}na}, {Lippa}, {Lenzen}, {Maire},
  {Molli{\`e}re}, {Ott}, {Paumard}, {Perraut}, {Perrin}, {Pueyo}, {Rabien},
  {Ram{\'\i}rez}, {Rau}, {Rodr{\'\i}guez-Coira}, {Rousset}, {Sanchez-Bermudez},
  {Scheithauer}, {Schuhler}, {Straub}, {Straubmeier}, {Sturm}, {Tacconi},
  {Vincent}, {van Dishoeck}, {von Fellenberg}, {Wank}, {Waisberg}, {Widmann},
  {Wieprecht}, {Wiest}, {Wiezorrek}, {Woillez}, {Yazici}, {Ziegler}, \&
  {Zins}}]{GRAVITY2019}
{Gravity Collaboration}, {Lacour}, S., {Nowak}, M., {et~al.} 2019, \aap, 623,
  L11

\bibitem[{{Gravity Collaboration} {et~al.}(2020){Gravity Collaboration},
  {Nowak}, {Lacour}, {Molli{\`e}re}, {Wang}, {Charnay}, {van Dishoeck},
  {Abuter}, {Amorim}, {Berger}, {Beust}, {Bonnefoy}, {Bonnet}, {Brandner},
  {Buron}, {Cantalloube}, {Collin}, {Chapron}, {Cl{\'e}net}, {Coud{\'e} Du
  Foresto}, {de Zeeuw}, {Dembet}, {Dexter}, {Duvert}, {Eckart}, {Eisenhauer},
  {F{\"o}rster Schreiber}, {F{\'e}dou}, {Garcia Lopez}, {Gao}, {Gendron},
  {Genzel}, {Gillessen}, {Hau{\ss}mann}, {Henning}, {Hippler}, {Hubert},
  {Jocou}, {Kervella}, {Lagrange}, {Lapeyr{\`e}re}, {Le Bouquin}, {L{\'e}na},
  {Maire}, {Ott}, {Paumard}, {Paladini}, {Perraut}, {Perrin}, {Pueyo}, {Pfuhl},
  {Rabien}, {Rau}, {Rodr{\'\i}guez-Coira}, {Rousset}, {Scheithauer},
  {Shangguan}, {Straub}, {Straubmeier}, {Sturm}, {Tacconi}, {Vincent},
  {Widmann}, {Wieprecht}, {Wiezorrek}, {Woillez}, {Yazici}, \&
  {Ziegler}}]{GRAVITY2020}
{Gravity Collaboration}, {Nowak}, M., {Lacour}, S., {et~al.} 2020, \aap, 633,
  A110

\bibitem[{{Hasegawa} \& {Pudritz}(2010)}]{HP10}
{Hasegawa}, Y. \& {Pudritz}, R.~E. 2010, \apjl, 710, L167

\bibitem[{{Hasegawa} \& {Pudritz}(2011)}]{HP11}
{Hasegawa}, Y. \& {Pudritz}, R.~E. 2011, \mnras, 417, 1236

\bibitem[{{Hasegawa} \& {Pudritz}(2013)}]{HP13}
{Hasegawa}, Y. \& {Pudritz}, R.~E. 2013, \apj, 778, 78

\bibitem[{{Helled} {et~al.}(2014){Helled}, {Bodenheimer}, {Podolak}, {Boley},
  {Meru}, {Nayakshin}, {Fortney}, {Mayer}, {Alibert}, \& {Boss}}]{Helled2014}
{Helled}, R., {Bodenheimer}, P., {Podolak}, M., {et~al.} 2014, in Protostars
  and Planets VI, ed. H.~{Beuther}, R.~S. {Klessen}, C.~P. {Dullemond}, \&
  T.~{Henning}, 643

\bibitem[{{Helling} {et~al.}(2014){Helling}, {Woitke}, {Rimmer}, {Kamp}, {Thi},
  \& {Meijerink}}]{Helling14}
{Helling}, C., {Woitke}, P., {Rimmer}, P.~B., {et~al.} 2014, Life, 4
  [\eprint[arXiv]{1403.4420}]

\bibitem[{{Ida} \& {Lin}(2004)}]{IL04a}
{Ida}, S. \& {Lin}, D.~N.~C. 2004, \apj, 604, 388

\bibitem[{{Ikoma} {et~al.}(2000){Ikoma}, {Nakazawa}, \& {Emori}}]{Ikoma2000}
{Ikoma}, M., {Nakazawa}, K., \& {Emori}, H. 2000, \apj, 537, 1013

\bibitem[{{Johansen} {et~al.}(2007){Johansen}, {Oishi}, {Mac Low}, {Klahr},
  {Henning}, \& {Youdin}}]{Johansen2007}
{Johansen}, A., {Oishi}, J.~S., {Mac Low}, M.-M., {et~al.} 2007, \nat, 448,
  1022

\bibitem[{{Klarmann} {et~al.}(2018){Klarmann}, {Ormel}, \&
  {Dominik}}]{Klarmann2018}
{Klarmann}, L., {Ormel}, C.~W., \& {Dominik}, C. 2018, ArXiv e-prints
  [\eprint[arXiv]{1809.01648}]

\bibitem[{{Kokubo} \& {Ida}(2002)}]{KI02}
{Kokubo}, E. \& {Ida}, S. 2002, \apj, 581, 666

\bibitem[{{Kreidberg} {et~al.}(2014){Kreidberg}, {Bean}, {D{\'e}sert}, {Line},
  {Fortney}, {Madhusudhan}, {Stevenson}, {Showman}, {Charbonneau},
  {McCullough}, {Seager}, {Burrows}, {Henry}, {Williamson}, {Kataria}, \&
  {Homeier}}]{Kre14}
{Kreidberg}, L., {Bean}, J.~L., {D{\'e}sert}, J.-M., {et~al.} 2014, \apjl, 793,
  L27

\bibitem[{{Krijt} {et~al.}(2020){Krijt}, {Bosman}, {Zhang}, {Schwarz},
  {Ciesla}, \& {Bergin}}]{Krijt2020}
{Krijt}, S., {Bosman}, A.~D., {Zhang}, K., {et~al.} 2020, arXiv e-prints,
  arXiv:2007.09517

\bibitem[{{Lambrechts} \& {Johansen}(2014)}]{LambJoh2014}
{Lambrechts}, M. \& {Johansen}, A. 2014, \aap, 572, A107

\bibitem[{{Lee} {et~al.}(2010){Lee}, {Bergin}, \& {Nomura}}]{Lee2010}
{Lee}, J.-E., {Bergin}, E.~A., \& {Nomura}, H. 2010, \apjl, 710, L21

\bibitem[{{Li} {et~al.}(2020){Li}, {Ingersoll}, {Bolton}, {Levin}, {Janssen},
  {Atreya}, {Lunine}, {Steffes}, {Brown}, {Guillot}, {Allison}, {Arballo},
  {Bellotti}, {Adumitroaie}, {Gulkis}, {Hodges}, {Li}, {Misra}, {Orton},
  {Oyafuso}, {Santos-Costa}, {Waite}, \& {Zhang}}]{Li2020}
{Li}, C., {Ingersoll}, A., {Bolton}, S., {et~al.} 2020, Nature Astronomy

\bibitem[{{Lin} \& {Papaloizou}(1986)}]{LP86}
{Lin}, D.~N.~C. \& {Papaloizou}, J. 1986, \apj, 309, 846

\bibitem[{{Line} {et~al.}(2014){Line}, {Knutson}, {Wolf}, \& {Yung}}]{Line2014}
{Line}, M.~R., {Knutson}, H., {Wolf}, A.~S., \& {Yung}, Y.~L. 2014, \apj, 783,
  70

\bibitem[{{Lyra} {et~al.}(2010){Lyra}, {Paardekooper}, \& {Mac Low}}]{Lyra2010}
{Lyra}, W., {Paardekooper}, S.-J., \& {Mac Low}, M.-M. 2010, \apjl, 715, L68

\bibitem[{{MacDonald} \& {Madhusudhan}(2019)}]{MacDonald2019}
{MacDonald}, R.~J. \& {Madhusudhan}, N. 2019, arXiv e-prints
  [\eprint[arXiv]{1903.09151}]

\bibitem[{{Madhusudhan}(2012)}]{Madu2012}
{Madhusudhan}, N. 2012, \apj, 758, 36

\bibitem[{{Madhusudhan} {et~al.}(2014){Madhusudhan}, {Amin}, \&
  {Kennedy}}]{Madu2014}
{Madhusudhan}, N., {Amin}, M.~A., \& {Kennedy}, G.~M. 2014, \apjl, 794, L12

\bibitem[{{Manara} {et~al.}(2018){Manara}, {Morbidelli}, \&
  {Guillot}}]{Manara2018}
{Manara}, C.~F., {Morbidelli}, A., \& {Guillot}, T. 2018, \aap, 618, L3

\bibitem[{{Masset} {et~al.}(2006){Masset}, {Morbidelli}, {Crida}, \&
  {Ferreira}}]{Masset06}
{Masset}, F.~S., {Morbidelli}, A., {Crida}, A., \& {Ferreira}, J. 2006, \apj,
  642, 478

\bibitem[{{McNally} {et~al.}(2018){McNally}, {Nelson}, \&
  {Paardekooper}}]{McNally2018}
{McNally}, C.~P., {Nelson}, R.~P., \& {Paardekooper}, S.-J. 2018, \mnras, 477,
  4596

\bibitem[{{McNally} {et~al.}(2020){McNally}, {Nelson}, {Paardekooper},
  {Ben{\'\i}tez-Llambay}, \& {Gressel}}]{McNally2020}
{McNally}, C.~P., {Nelson}, R.~P., {Paardekooper}, S.-J.,
  {Ben{\'\i}tez-Llambay}, P., \& {Gressel}, O. 2020, \mnras
  [\eprint[arXiv]{2002.11161}]

\bibitem[{{Mordasini} {et~al.}(2009){Mordasini}, {Alibert}, {Benz}, \&
  {Naef}}]{Mordasini2009b}
{Mordasini}, C., {Alibert}, Y., {Benz}, W., \& {Naef}, D. 2009, \aap, 501, 1161

\bibitem[{{Mordasini} {et~al.}(2015){Mordasini}, {Molli{\`e}re}, {Dittkrist},
  {Jin}, \& {Alibert}}]{Mordasini15}
{Mordasini}, C., {Molli{\`e}re}, P., {Dittkrist}, K.-M., {Jin}, S., \&
  {Alibert}, Y. 2015, International Journal of Astrobiology, 14, 201

\bibitem[{{Mordasini} {et~al.}(2016){Mordasini}, {van Boekel}, {Molli{\`e}re},
  {Henning}, \& {Benneke}}]{Mordasini16}
{Mordasini}, C., {van Boekel}, R., {Molli{\`e}re}, P., {Henning}, T., \&
  {Benneke}, B. 2016, \apj, 832, 41

\bibitem[{{Morley} {et~al.}(2017){Morley}, {Knutson}, {Line}, {Fortney},
  {Thorngren}, {Marley}, {Teal}, \& {Lupu}}]{Morley2017}
{Morley}, C.~V., {Knutson}, H., {Line}, M., {et~al.} 2017, \aj, 153, 86

\bibitem[{{Moses} {et~al.}(2013){Moses}, {Madhusudhan}, {Visscher}, \&
  {Freedman}}]{Moses13}
{Moses}, J.~I., {Madhusudhan}, N., {Visscher}, C., \& {Freedman}, R.~S. 2013,
  \apj, 763, 25

\bibitem[{{{\"O}berg} {et~al.}(2011){{\"O}berg}, {Murray-Clay}, \&
  {Bergin}}]{Oberg11}
{{\"O}berg}, K.~I., {Murray-Clay}, R., \& {Bergin}, E.~A. 2011, \apjl, 743, L16

\bibitem[{{{\"O}berg} \& {Wordsworth}(2019)}]{Oberg2019}
{{\"O}berg}, K.~I. \& {Wordsworth}, R. 2019, \aj, 158, 194

\bibitem[{{Ormel} \& {Klahr}(2010)}]{Ormel2010}
{Ormel}, C.~W. \& {Klahr}, H.~H. 2010, \aap, 520, A43

\bibitem[{{Pinhas} {et~al.}(2019){Pinhas}, {Madhusudhan}, {Gandhi}, \&
  {MacDonald}}]{Pinhas2019}
{Pinhas}, A., {Madhusudhan}, N., {Gandhi}, S., \& {MacDonald}, R. 2019, \mnras,
  482, 1485

\bibitem[{{Pinilla} {et~al.}(2016){Pinilla}, {Klarmann}, {Birnstiel},
  {Benisty}, {Dominik}, \& {Dullemond}}]{Pinilla2016}
{Pinilla}, P., {Klarmann}, L., {Birnstiel}, T., {et~al.} 2016, \aap, 585, A35

\bibitem[{{Pirani} {et~al.}(2019){Pirani}, {Johansen}, {Bitsch}, {Mustill}, \&
  {Turrini}}]{Pirani2019}
{Pirani}, S., {Johansen}, A., {Bitsch}, B., {Mustill}, A.~J., \& {Turrini}, D.
  2019, \aap, 623, A169

\bibitem[{{Pollack} {et~al.}(1996){Pollack}, {Hubickyj}, {Bodenheimer},
  {Lissauer}, {Podolak}, \& {Greenzweig}}]{Pollack1996}
{Pollack}, J.~B., {Hubickyj}, O., {Bodenheimer}, P., {et~al.} 1996, \icarus,
  124, 62

\bibitem[{{Pontoppidan} {et~al.}(2014){Pontoppidan}, {Salyk}, {Bergin},
  {Brittain}, {Marty}, {Mousis}, \& {{\"O}berg}}]{Pon14}
{Pontoppidan}, K.~M., {Salyk}, C., {Bergin}, E.~A., {et~al.} 2014, Protostars
  and Planets VI, 363

\bibitem[{{Raymond} \& {Morbidelli}(2014)}]{Raymond2014}
{Raymond}, S.~N. \& {Morbidelli}, A. 2014, in IAU Symposium, Vol. 310, Complex
  Planetary Systems, Proceedings of the International Astronomical Union,
  194--203

\bibitem[{{Tamayo} {et~al.}(2015){Tamayo}, {Triaud}, {Menou}, \&
  {Rein}}]{Tamayo15}
{Tamayo}, D., {Triaud}, A.~H.~M.~J., {Menou}, K., \& {Rein}, H. 2015, \apj,
  805, 100

\bibitem[{{Tychoniec} {et~al.}(2020){Tychoniec}, {Manara}, {Rosotti}, {van
  Dishoeck}, {Cridland }, {Hsieh}, {Murillo}, {Segura-Cox}, {van Terwisga}, \&
  {Tobin}}]{Tychoniec2020}
{Tychoniec}, {\L}., {Manara}, C.~F., {Rosotti}, G.~P., {et~al.} 2020, arXiv
  e-prints, arXiv:2006.02812

\bibitem[{{Tychoniec} {et~al.}(2018){Tychoniec}, {Tobin}, {Karska}, {Chandler},
  {Dunham}, {Harris}, {Kratter}, {Li}, {Looney}, {Melis}, {P{\'e}rez},
  {Sadavoy}, {Segura-Cox}, \& {van Dishoeck}}]{Tychoniec2018}
{Tychoniec}, {\L}., {Tobin}, J.~J., {Karska}, A., {et~al.} 2018, \apjs, 238, 19

\bibitem[{{Voelkel} {et~al.}(2020){Voelkel}, {Klahr}, {Mordasini},
  {Emsenhuber}, \& {Lenz}}]{Voelkel2020}
{Voelkel}, O., {Klahr}, H., {Mordasini}, C., {Emsenhuber}, A., \& {Lenz}, C.
  2020, arXiv e-prints, arXiv:2004.03492

\bibitem[{{Walsh} {et~al.}(2015){Walsh}, {Nomura}, \& {van Dishoeck}}]{Walsh15}
{Walsh}, C., {Nomura}, H., \& {van Dishoeck}, E. 2015, \aap, 582, A88

\bibitem[{{Walsh} {et~al.}(2011){Walsh}, {Morbidelli}, {Raymond}, {O'Brien}, \&
  {Mandell}}]{Walsh2011}
{Walsh}, K.~J., {Morbidelli}, A., {Raymond}, S.~N., {O'Brien}, D.~P., \&
  {Mandell}, A.~M. 2011, \nat, 475, 206

\bibitem[{{Ward}(1991)}]{W91}
{Ward}, W.~R. 1991, in Lunar and Planetary Inst.~Technical Report, Vol.~22,
  Lunar and Planetary Science Conference

\bibitem[{{Ward}(1997)}]{Ward1997}
{Ward}, W.~R. 1997, \apjl, 482, L211

\bibitem[{{Wei} {et~al.}(2019){Wei}, {Nomura}, {Lee}, {Ip}, {Walsh}, \&
  {Millar}}]{Wei2019}
{Wei}, C.-E., {Nomura}, H., {Lee}, J.-E., {et~al.} 2019, \apj, 870, 129

\bibitem[{{Wittenmyer} {et~al.}(2011){Wittenmyer}, {Tinney}, {O'Toole},
  {Jones}, {Butler}, {Carter}, \& {Bailey}}]{Wittenmyer2011}
{Wittenmyer}, R.~A., {Tinney}, C.~G., {O'Toole}, S.~J., {et~al.} 2011, \apj,
  727, 102

\bibitem[{{Wright} {et~al.}(2012){Wright}, {Marcy}, {Howard}, {Johnson},
  {Morton}, \& {Fischer}}]{Wright2012}
{Wright}, J.~T., {Marcy}, G.~W., {Howard}, A.~W., {et~al.} 2012, \apj, 753, 160

\end{thebibliography}

\appendix

\section{Breakdown of most abundant species}\label{sec:species}

\begin{figure*}
    \centering
    \includegraphics[width=\textwidth]{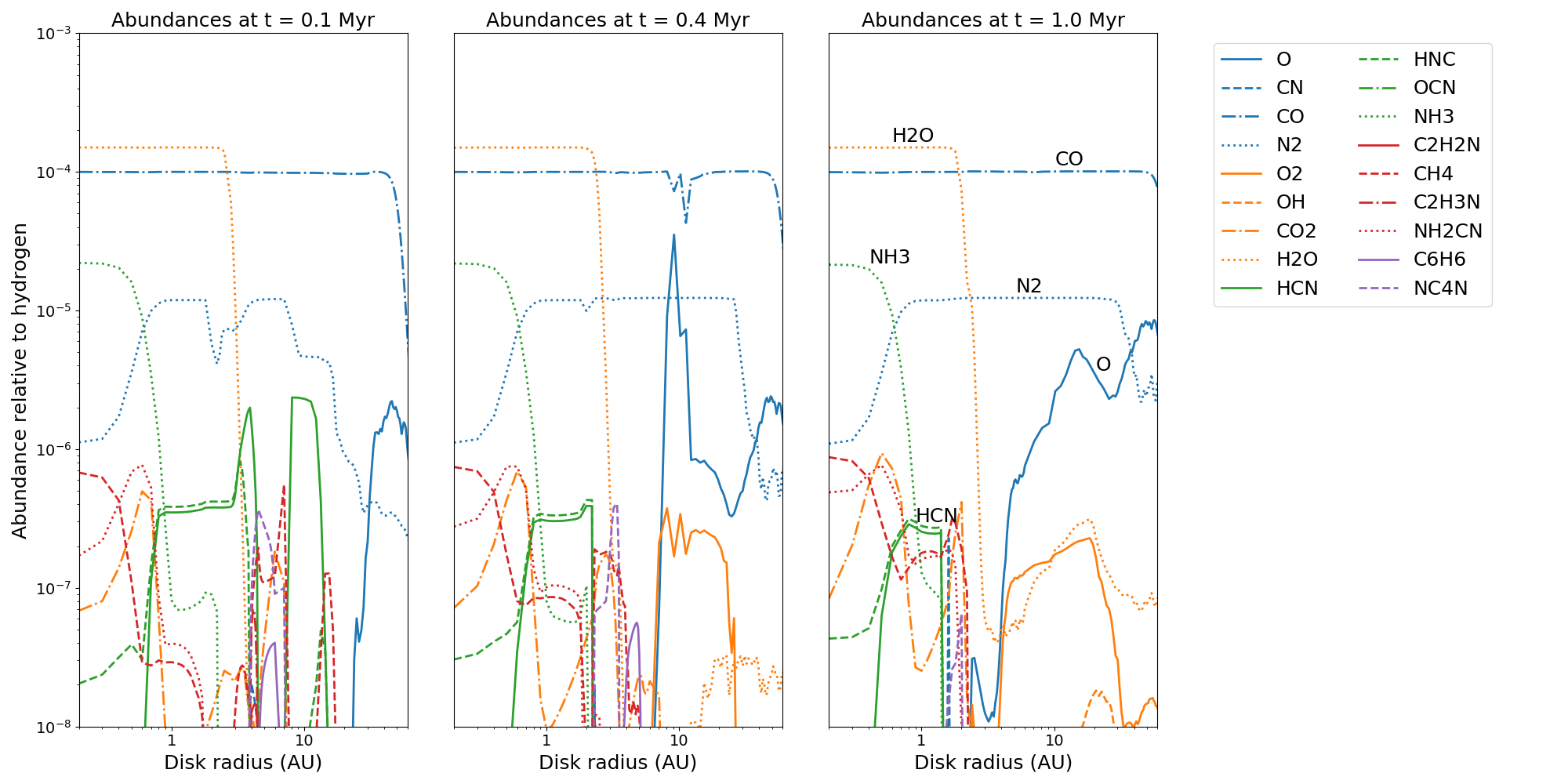}
    \caption{Abundances for the most abundant gas species in the chemical model. Annotated on the right-most panel are the most abundant species in our chemical model.}
    \label{fig:chem_test01}
\end{figure*}

\begin{figure*}
    \centering
    \includegraphics[width=\textwidth]{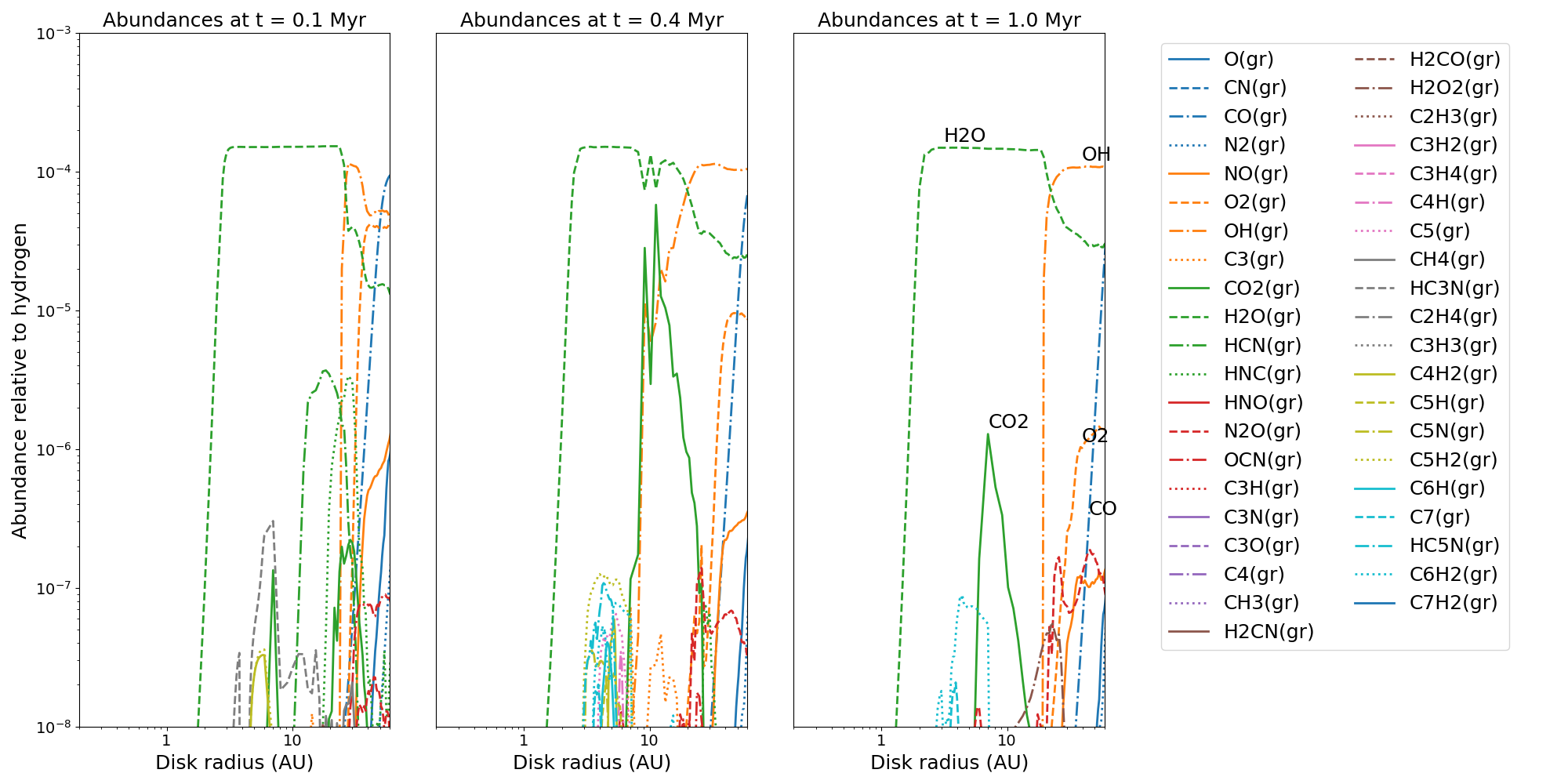}
    \caption{Same as Figure \ref{fig:chem_test01} but for the most abundant ice species.}
    \label{fig:chem_test02}
\end{figure*}

Here we show the most abundant species found in one particular chemical model in our population of disks. The carbon and oxygen carrying species are predominately \ce{H2O}, OH, and CO. there are periods of time (and ranges of radii) where \ce{CO2} becomes abundant on the icy grains, while molecular oxygen and atomic oxygen become abundant in the gas. Frozen OH is mainly made from the dissociation of frozen water by UV photons induced by collisions of cosmic rays with molecular hydrogen. A very small amount of carbon and nitrogen bearing species like \ce{CH4}, HCN, and \ce{C6H6} can be found in the gas phase, however these species do not survive throughout the chemical evolution of the disk. Unlike some chemical models, we do not produce large amounts of gaseous \ce{CO2} - which is generally produced through grain surface reactions that are not included in our model.

\end{document}